\documentclass[a4paper,11pt]{article}

\usepackage{amsmath,amsfonts}
\usepackage{bm}
\usepackage{typearea}
\usepackage{authblk}
\usepackage{float}
\typearea{12}

\usepackage{graphicx}

\makeatletter
\@addtoreset{equation}{section}
\makeatother

\def\diag{\mathop{\rm diag}\nolimits}

\begin{document}

\title{Pricing Bermudan Swaption under Two Factor Hull-White Model with Fast Gauss Transform}
\author[1,2]{Tomohisa Yamakami}
\author[1]{Yuki Takeuchi}

\affil[1]{Mizuho-DL Financial Technology Co., Ltd.\footnote{The opinions expressed herein are only those of the authors. They do not represent the official views of the Mizuho-DL Financial Technology Co., Ltd.}}
\affil[2]{Graduate School of Economics, The University of Tokyo}

\date{\today}
\maketitle

\begin{abstract}
  This paper describes a fast and stable algorithm for evaluating Bermudan swaption under the two factor Hull-White model.
  We discretize the calculation of the expected value in the evaluation of Bermudan swaption by numerical integration, and Gaussian kernel sums appears in it.
  The fast Gauss transform can be applied to these Gaussian kernel sums, and it reduces computational complexity from $O(N^2)$ to $O(N)$ as grid points number $N$ of numerical integration.
  We also propose to stabilize the computation under the condition that the correlation is close to $-1$ by introducing the grid rotation.
  Numerical experiments using actual market data show that our method reduces the computation time significantly compared to the method without the fast Gauss transform.
  They also show that the method of the grid rotation contributes to computational stability in the situations where the correlation is close to $-1$ and time step is short.
\end{abstract}

\section{Introduction}

Bermudan swaption is an option which has several exercise days and enters an interest rate swap when it is exercised.
This instrument is widely traded for the purpose of structuring bonds and loans which have early redemption conditions.
Evaluating Bermudan swaptions is an important part of financial businesses since traders have to quote and manage portfolios which contain Bermudan swaptions.

According to the risk-neutral valuation, the price of a derivative can be calculated based on the expectation of discounted value of the payoff.
If the payoff, model, and other conditions are simple, there are analytic solutions. Otherwise, numerical methods are often required to calculate the expected value.
Considering a valuation of a Bermudan swaption, most situations require complex calculations since a valuation of Bermudan swaption needs
calculations of both expectation of payoff and optimal exercise time.
Using dynamic programming, we can derive a procedure that the exercise decision is made based on the large-small relation of exercised swap value and
holding value for future exercise rights. However, the holding value itself has a recursive structure that depends on the exercise value and the holding value at the time of the next exercise.
Due to this structure, most models cannot provide any analytical solution of Bermudan swaption, and numerical calculations are required.
In practice, many iterative calculations are required not only for price calculations
but also for calculations of sensitivities for risk management.
Therefore, to speed up these numerical calculations is an important issue.

Several methods are known for the valuation of Bermudan-type derivatives.
Longstaff and Schwartz \cite{longstaff2001valuing} proposed least square Monte Carlo (LSMC)
which is a general-purpose method that can be used to evaluate derivatives including early exercises.
Karlsson et al. \cite{feng2016efficient} apply the method to calculating the Bermudan swaption.
These methods have a slow convergence issue of the Monte Carlo method.
The calculation errors of these methods are reduced as order $O(N^{-0.5})$ with path number $N$,
which means that we need $100$ times paths to reduce the error to $1/10$.
Hull and White \cite{hull1990valuing}, Li et al. \cite{li1995lattice}, and Lee and Yang \cite{lee2020finite}
proposed Tree and finite difference methods solving partial differential equations
which require a finer grid in the time direction as well as in the state space direction to improve the accuracy of the calculation,
and they require more computations than the product of the number of discretization grids of state space and time directions.
Load et al. \cite{lord2008fast} and Karlsson et al. \cite{feng2016efficient} introduced the CONV method which uses the fast Fourier transform.
It can be calculated in $(O(N\log N))$ complexity for the number of spatial discretization $N$
if the transition density function or its characteristic function is known.
The fast Gauss transform (FGT) introduced by Greengard and Strain \cite{greengard1991fast} is a method for calculating the weighted sums of a Gaussian function at multiple points,
whose computational complexity is $O(N+N')$ instead of $O(NN')$ by direct calculation, where the number of input grids is $N$ and the number of output grids is $N'$.
Yang et al. \cite{yang2003improved} extended this method to any dimension and reducing computational complexity as increasing dimension.
The weighted sums of Gaussian functions often appear in the calculation of derivatives, and in fact,
Broadie and Yamamoto\cite{broadie2003application} showed that the FGT can be applied to the evaluation of Bermudan-type and other derivatives.

After the financial crisis, credit valuation adjustments (CVA and widely XVA) have become essential for pricing and risk management.
In CVA calculation which treats counter party credit risk of derivatives, it is necessary to evaluate transactions involving various assets in the netting set at the same time,
so hybrid models incorporating various factors such as interest rates, exchange rates, stocks, and default probabilities are used.
In addition, the calculation of exposures at many points in time and in many model state spaces is needed, which makes the calculation time of the models assuming default much longer than those assuming default-free.
In these situations, Gaussian models including the one or two factor Hull-White model are being re-evaluated as an interest rate model for CVA.
There are many reasons such as its ability to perfectly reproduce the yield curve, its ability to obtain many analytical solutions for various interest rate products,
its ease of handling interest rates as a Gaussian distribution even when constructing a hybrid model with exchange rates,
and its suitability for the recent negative interest rate environment.

The parameters of the two factor Hull-White model (equivalent to G2++ model) are calibrated to fit the cap and swaption volatilities of the market.
Brigo and Mercurio\cite{brigo2001interest} have mentioned that the correlation parameter of the model is sometimes very close to $-1$, especially when fitting to Cap volatilities.
In practice, the correlation is sometimes very close to $-1$ even if we are fitting model to Swaption volatilities.
Stability of the calculation in such situations is essential. Even if the situation occurs only 1\%, it occurs an average of 2.5 times per year in daily calculation.

In this paper, we consider the evaluation of Bermudan swaption by the two factor Hull-White model,
and discuss an efficient and stable calculation method using the fast Gauss transform and grid rotation for the integral calculation appearing in the valuation formula.
Our method is not limited to the evaluation of Bermudan swaptions but can be applied to the evaluation of a wider range of products.
Also, it is applicable wide range of models whose transition density is a multivariate Gaussian distribution of arbitrary dimensions.

This paper is organized as follows.
In Section 2, we outline the evaluation method of interest rate derivatives using the two factor Hull-White model, the multi-curve modeling, the evaluation formula of Bermudan swaption, and the principle of the fast Gauss transform.
In Section 3, we construct an approximation by numerical integration of the integrals appearing in the evaluation formula of Bermudan swaption based on the two factor Hull-White model
and show that the FGT can be applied. We also explain grid rotation technique as stabilizing calculation in the situation that correlation is close to $-1$.
In Section 4, numerical examples are presented. We conclude in Section 5.

\section{Bermudan Swaption Pricing on Two Factor Hull-White Model}

\subsection{Two Factor Hull-White Model}
We assume the ordinary filtered probability space $(\Omega,\mathcal{F},\mathbb{P};\left\{\mathcal{F}_{t}\right\}_{t\geq0})$
Let $R_{t}$ be the spot interest rate and we consider prices of derivatives in time $[0,T_{h}]$.
If the market is arbitrage-free and complete, there exists an equivalence martingale measure $\mathbb{Q}$ such that the relative prices of any derivative to the money market account,
$\exp\left\{\int_{0}^{t}R_{s}ds\right\}$, become a martingale and it is called the risk-neutral measure.
Hereafter, we use $E[\cdot]$ as the expectation under the measure $\mathbb{Q}$ and
introduce the conditional expectation $E_{t}[\cdot]=E[\cdot|\mathcal{F}_{t}]$ for brevity.

Let $\bm{X}_{t}=(X_{0,t}\ X_{1,t})^{T}$ be a two-dimensional stochastic process and
$\bm{W}_{t}=(W_{0,t}\ W_{1,t})^{T}$ be a $\left\{\mathcal{F}_{t}\right\}_{t\geq0}$-adapted two-dimensional independent standard Brownian motion under a risk-neutral measure $\mathbb{Q}$.
The two factor Hull-White model which is an interest rate model for the $R_{t}$ is defined by
\begin{eqnarray}
  R_{t}&=&\bm{1}^{T}\bm{X}_{t}+\psi(t),\nonumber\\
  d \bm{X}_{t}&=&-\bm{\kappa}(t)\bm{X}_{t}dt+\bm{\sigma}(t)\bm{C}_{\rho}d\bm{W}_{t},\label{formula:G2PPDef}
\end{eqnarray}
with $\bm{1}=(1\ 1)^{T}$ and the variables in \eqref{formula:G2PPDef} are
\begin{eqnarray*}
  \bm{\kappa}(t)&=&\diag{(\kappa_{0}(t), \kappa_{1}(t))},\\
  \bm{\sigma}(t)&=&\diag{(\sigma_{0}(t), \sigma_{1}(t))},\\
  \bm{C}_{\rho}&=&
\begin{pmatrix}
  1 & 0\\
  \rho & \sqrt{1-\rho^{2}}
\end{pmatrix},
\end{eqnarray*}
where $\diag{(\cdot)}$ is a diagonal matrix whose arguments are diagonal elements and $\psi(t)$, $\bm{\kappa}(t)$, $\bm{\sigma}(t)$, and $\rho$ are deterministic.
Integrating (\ref{formula:G2PPDef}) gives
\begin{eqnarray}
  \bm{X}_{\tau}&=&\bm{\mu}(t,\tau)\bm{X}_{t}+\bm{U}(t,\tau),\label{formula:G2PPInt1}\\
  \int_{t}^{\tau}\bm{X}_{s}ds&=&\bm{\nu}(t,\tau)\bm{X}_{t}+\bm{V}(t,\tau).\label{formula:G2PPInt2}
\end{eqnarray}
In this paper, integrals of vectors and matrices are denoted as vectors and matrices with integrals of each element.
The variables in (\ref{formula:G2PPInt1}) and (\ref{formula:G2PPInt2}) are written by
\begin{eqnarray*}
  \bm{\mu}(t,\tau)&=&
  \diag{(\mu_{0}(t,\tau), \mu_{1}(t,\tau))}=\exp\left\{-\int_{t}^{\tau}\bm{\kappa}(s)ds\right\},\\
  \bm{\nu}(t,\tau)&=&
  \diag{(\nu_{0}(t,\tau), \nu_{1}(t,\tau))}=\int_{t}^{\tau}\bm{\mu}(t,s)ds,\\
  \bm{U}(t,\tau)&=&
  \begin{pmatrix}
    U_{0}(t,\tau)\\
    U_{1}(t,\tau)
  \end{pmatrix}=\int_{t}^{\tau}\bm{\mu}(s,\tau)\bm{\sigma}(s)\bm{C}_{\rho}d\bm{W}_{s},\\
  \bm{V}(t,\tau)&=&
  \begin{pmatrix}
    V_{0}(t,\tau)\\
    V_{1}(t,\tau)
  \end{pmatrix}=\int_{t}^{\tau}\bm{\nu}(s,\tau)\bm{\sigma}(s)\bm{C}_{\rho}d\bm{W}_{s},
\end{eqnarray*}
and the conditional variance-covariance matrix of $\bm{U}(t,\tau)$ under $\sigma$-algebra $\mathcal{F}_{t}$ is calculated as
\begin{eqnarray*}
  \bm{\Sigma}(t,\tau)&=&\mathbb{E}_{t}\left[\bm{U}(t,\tau)\bm{U}^{T}(t,\tau)\right]\nonumber\\
  &=&\int_{t}^{\tau}\bm{\mu}(s,\tau)\bm{\sigma}(s)\bm{\Lambda}_{\rho}\bm{\sigma}^{T}(s)\bm{\mu}^{T}(s,\tau).
\end{eqnarray*}

The discounted bond price $P_{t}(\tau)$ with maturity $\tau$ at evaluation time $t$ is written by
\begin{eqnarray*}
  P_{t}(\tau)&=&\mathbb{E}_{t}\left[e^{-\int_{t}^{\tau}R_{s}ds}\right]\nonumber\\
  &=&F(t,\tau)e^{-\bm{1}^{T}\bm{\nu}(t,\tau)\bm{X}_{t}},\\
  F(t,\tau)&=&e^{\frac{1}{2}v(t,\tau)-\int_{t}^{\tau}\psi(s)ds},\\
  v(t,T)&=&\int_{t}^{\tau}\bm{1}^{T}\bm{\nu}(s,\tau)\bm{\sigma}(s)\bm{\Lambda}_{\rho}\bm{\sigma}^{T}(s)\bm{\nu}^{T}(s,\tau)\bm{1}ds,
\end{eqnarray*}
Therefore, if we define $\psi(s)$ as
\begin{eqnarray*}
  \int_{0}^{\tau}\psi(s)ds&=&\frac{1}{2}v(0,\tau)-\log P_{0}(\tau),
\end{eqnarray*}
this model can reproduce the discount factor at time 0 perfectly.

Next, we derive the expression under the forward risk neutral measure.
Let the Radon-Nikodym derivative process be
\begin{eqnarray*}
  \mathbb{E}_{t}\left[\frac{d\mathbb{Q}^{\tau}}{d\mathbb{Q}}\right]&=&\mathbb{E}_{t}\left[\frac{1}{e^{\int_{0}^{\tau}R_{s}ds}P_{0}(\tau)}\right]\nonumber\\
  &=&e^{-\frac{1}{2}\left(v(0,\tau)-v(t,\tau)\right)-\int_{0}^{t}\bm{1}^{T}\bm{\nu}(s,\tau)\bm{\sigma}(s)\bm{C}_{\rho}d\bm{W}_{s}}.
\end{eqnarray*}
From Girsanov's theorem, Brownian motion defined by
\begin{eqnarray*}
  d\bm{W}_{t}^{\tau}&=&
  \begin{pmatrix}
    dW_{0,t}^{\tau}\\
    dW_{1,t}^{\tau}
  \end{pmatrix}
  =
  d\bm{W}_{t}+\bm{C}_{\rho}^{T}\bm{\sigma}(t)\bm{\nu}(t,\tau)\bm{1}dt,
  \end{eqnarray*}
become an independent standard Brownian motion under the measure $\mathbb{Q}^{\tau}$ which is called $\tau$ forward risk neutral measure.
Under $\mathbb{Q}^{\tau}$, the stochastic process of $\bm{X}_{t}$ can be written as
\begin{eqnarray}
  d \bm{X}_{t}&=&
  \left\{\bm{\theta}(t)-\bm{\kappa}(t)\bm{X}_{t}\right\}dt
  +
  \bm{\sigma}(t)
  \bm{C}_{\rho}
  d\bm{W}_{t}^{\tau},\label{formula:FwddX}\\
  \bm{\theta}(t)&=&-\bm{\sigma}(t)\bm{\Lambda}_{\rho}\bm{\sigma}(t)\bm{\nu}(t,\tau)\bm{1},\nonumber
\end{eqnarray}
Integrating (\ref{formula:FwddX}) gives
\begin{eqnarray}
  \bm{X}_{\tau}&=&\bm{\mu}(t,\tau)\bm{X}_{t}+\bm{\eta}(t,\tau)+\bm{U}_{0}^{\tau}(t,\tau)\label{formula:ForwardX}
\end{eqnarray}
where variables are written by
\begin{eqnarray*}
  \bm{\eta}(t,\tau)&=&
  \begin{pmatrix}
    \eta_{0}(t,\tau)\\
    \eta_{1}(t,\tau)
  \end{pmatrix}=\int_{t}^{\tau}\bm{\mu}(s,\tau)\bm{\theta}(s)ds,\\
  \bm{U}^{\tau}(t,\tau)&=&
  \begin{pmatrix}
    U_{0}^{\tau}(t,\tau)\\
    U_{1}^{\tau}(t,\tau)
  \end{pmatrix}=\int_{t}^{\tau}\bm{\mu}(s,\tau)\bm{\sigma}(s)\bm{C}_{\rho}d\bm{W}_{s}^{\tau},
\end{eqnarray*}
Note that the variance-covariance matrices of $\bm{U}^{\tau}(t,\tau)$ are the same as $\bm{U}(t,\tau)$.

Let $V(t,\bm{X}_{t})$ be derivative value at time $t$ and state $\bm{X}_{t}$.
If there are no cashflow between $[t,\tau]$, the relation between $V(t,\bm{X}_{t})$ and $V(\tau,\bm{X}_{\tau})$ is written by
\begin{eqnarray}
  V(t,\bm{X}_{t})&=&\mathbb{E}_{t}\left[e^{-\int_{t}^{\tau}R_{s}ds}V(\tau,\bm{X}_{\tau})\right]\nonumber\\
  &=&P_{t}(\tau)\mathbb{E}_{t}^{\tau}\left[V(\tau,\bm{X}_{\tau})\right].\label{formula:G2Evaluation}
\end{eqnarray}

In the two factor Hull-White model with constant coefficients, analytical solutions for cap and swaption prices are known (see e.g., \cite{brigo2001interest})
and the parameters $\bm{\kappa}(t),\bm{\sigma}(t),\rho$ are calibrated to market cap and swaption prices.
To evaluate a Bermudan swaption,
we sometimes calibrate parameters with co-terminal swaptions which satisfy that
the sums of the option maturity and the underlying swap period 
are equal to term of the Bermudan swaption, because trader often hedge Bermudan swaption with these co-terminal swaptions.

Brigo and Mercurio \cite{brigo2001interest} also mentioned that the correlation $\rho$ is often very close to $-1$ in the calibration to market Cap.
In practice, the correlation is sometimes very close to $-1$ even when we calibrate to swaptions.
Such conditions are prone to cause errors in numerical calculations and often require ingenuity.

\subsection{Multi Curve Model}
After the financial crisis, the theory of discounting has changed drastically as counterparty default risk has to be considered.
It has become necessary to use multiple curves (see e.g. \cite{RePEc:cfi:fseres:cf154}), such as using LIBOR, which will soon disappear by LIBOR reform and risk-free rate (RFR) will be standard, for a reference rate while using OIS for discounting collateralized derivatives.
This subsecsion introduces multi curve modeling with the assumption that the difference between the instantaneous forward rates of the risk-free rate and reference interest rate curves is deterministic.

Assuming $t\leq t_{s}<t_{e}$, we define forward-looking reference rate whose term is $[t_{s},t_{e}]$ and which is observed at $t$ as 
\begin{eqnarray}
  L_{t}(t_{s},t_{e})&=&\frac{1}{t_{e}-t_{s}}\left(\frac{P_{t}^{L}(t_{s})}{P_{t}^{L}(t_{e})}-1\right),\label{formula:multicurve_simplerate}\\
  P_{t}^{L}(\tau)&=&\exp\left\{\int_{t}^{\tau}\psi(s)-\psi^{L}(s)ds\right\}P_{t}(\tau).\nonumber
\end{eqnarray}
At this time, the value of the floating cashflow with reference rate $L_{t_{s}}(t_{s},t_{e})$ is
\begin{eqnarray}
  \mathbb{E}_{t}\left[L_{t_{s}}(t_{s},t_{e})e^{-\int_{t}^{t_{e}}R_{s}ds}\right]&=&\mathbb{E}_{t}\left[\frac{1}{t_{e}-t_{s}}\left(\frac{\mathbb{E}_{t_{s}}\left[e^{-\int_{t}^{t_{e}}R_{s}ds}\right]}{P_{t_{s}}^{L}(t_{e})}-e^{-\int_{t}^{t_{e}}R_{s}ds}\right)\right]\nonumber\\
  &=&\frac{1}{t_{e}-t_{s}}\left(e^{-\int_{t_{s}}^{t_{e}}\psi(s)-\psi^{L}(s)ds}P_{t}(t_{s})-P_{t}(t_{e})\right)\label{formula:multicurve_float_eq_df}\\
  &=&L_{t}(t_{s},t_{e})P_{t}(t_{e}).\nonumber
\end{eqnarray}
We find that $L_{t}(t_{s},t_{e})$ is a martingale under the measure $\mathbb{Q}^{t_{e}}$ and also is equivalent to a portfolio of discounted bonds.
Furthermore,
\begin{eqnarray*}
  \mathbb{E}_{t_{s}}^{t_{e}}\left[(1+(t_{m}-t_{s})L_{t_{s}}(t_{s},t_{m}))(1+(t_{e}-t_{m})L_{t_{m}}(t_{m},t_{e}))\right]&=&1+(t_{e}-t_{s})L_{t_{s}}(t_{s},t_{e}),
\end{eqnarray*}
is valid as $t_{s}<t_{m}<t_{e}$.

Let $t_{s}=t_{o,0}<\cdots<t_{o,N_{o}}=t_{e}$ be business days sequence and 
\begin{eqnarray*}
  L^{c}(t_{s},t_{e})=\frac{1}{t_{e}-t_{s}}\left[\prod_{i=0}^{N_{o}-1}\left\{1+(t_{o,i+1}-t_{o,i})L_{t_{o,i}}(t_{o,i},t_{o,i+1})\right\}-1\right],
\end{eqnarray*}
be compounding interest rate between $[t_{s},t_{e}]$,
we can derive the value of compounding rate cashflow as 
\begin{eqnarray}
  \mathbb{E}_{t}\left[L^{c}(t_{s},t_{e})e^{-\int_{t}^{t_{e}}R_{s}ds}\right]&=&L_{t}(t_{s},t_{e})P_{t}(t_{e}),\label{formula:RFR_in_arrier_value}
\end{eqnarray}
which equal to the value of forward looking reference rate.

(\ref{formula:multicurve_float_eq_df}) and (\ref{formula:RFR_in_arrier_value}) allows us to express the value of a swap as a portfolio of discounted bonds,
even if the discount and reference curves are different.

For example, let $t_{S,0}$ be the beginning of the swap period, and $t_{S,1},\cdots,t_{S,N_{S}}$ is the interest payment date of the swap,
the valuation $V_{S}(t)$ of a swap which receives a fixed interest rate $K$ with the same frequency on the fixed and floating legs is
\begin{eqnarray*}
  V_{S}(t)&=&\sum_{i=1}^{N_{S}}K(t_{S,i}-t_{S,i-1})P_{t}(t_{S,i})-\sum_{i=1}^{N_{S}}L_{t}(t_{S,i-1},t_{S,i})(t_{S,i}-t_{S,i-1})P_{t}(t_{S,i})\\
  &=&\sum_{i=1}^{N_{S}-1}\left\{K(t_{S,i}-t_{S,i-1})+e^{-\int_{t_{S,i-1}}^{t_{S,i}}\psi(s)-\psi^{L}(s)ds}-1\right\}P_{t}(t_{S,i})\\
  &&+\left\{1+K(t_{S,i}-t_{S,i-1})\right\}P_{t}(t_{S,N_{S}})-e^{-\int_{t_{S,0}}^{t_{S,1}}\psi(s)-\psi^{L}(s)ds}P_{t}(t_{S,0}).
\end{eqnarray*}
The analytical solution of swaption in the two factor Hull-White model is derived using the fact that this value at time $t_{S,0}$ is a monotonic function of the elements of the state variables (see e.g., \cite{brigo2001interest}).
In this multi-curve modeling, it may not be monotonic due to the presence of the coefficient $e^{-\int_{t_{s}}^{\tau}\psi(s)-\psi^{L}(s)ds}$.
We noted that there are some cases where analytical solution of swaption is not applicable.

\subsection{Bermudan Swaption}
A Bermudan swaption is an option which enters an interest rate swap if it is exercised
and there is a one-time opportunity to exercise the right from a predetermined set of dates.
In a normal transaction, an interest rate swap is defined as underlying asset and future cashflows from each exercise date are arise as exercising.
Here we introduce the recurrence relation for the Bermudan swaption without assuming any specific model.

Let $E(t_{1},\bm{X}_{t_{1}}),\cdots,E(t_{n},\bm{X}_{t_{n}})$ be the value of the swap entering by the exercise of the Bermudan swaption at
$\mathcal{T}=\{t_{1},\cdots,t_{n}\}$, where $\bm{X}_{t_{n}}$ is the state variable of the evaluation model at each time.

The swaps are commonly exchanges of fixed and floating interest rates, and equivalent to a portfolio of discounted bonds.
Caplet floorlets and other variable cashflows can also be included if their value can be written only in terms of the state variable $\bm{X}_{t_{n}}$.

The value of a Bermudan swaption which has not been exercised until time $t$ is described by
\begin{eqnarray*}
  V_{B}(t,X_{t})=\sup_{\bm{t}_{o}\in\mathcal{T}_{t}}\mathbb{E}_{t}\left[e^{-\int_{t}^{\bm{t}_{o}}R_{s}ds}\max\left\{E(\bm{t}_{o},\bm{X}_{\bm{t}_{o}}),0\right\}\right],
\end{eqnarray*}
with $\bm{t}_{o}$ as the stopping time taken among $\mathcal{T}_{t}=\mathcal{T}\cap[t,T_{h}]$.

$V_{B}(t,X_{t})$ is calculated by the following recurrent formula using dynamic programming with $k(t)$ as the smallest $k$ satisfying $t\leq t_{k}$.
\begin{eqnarray*}
  H_{n+1}(t,X_{t})&=&0,\\
  H_{k}(t,X_{t})&=&\mathbb{E}_{t}\left[e^{-\int_{t}^{t_{k}}R_{s}ds}\max\left\{E(t_{k},\bm{X}_{t_{k}}),H_{k+1}(t_{k},X_{t_{k}})\right\}\right] (t\leq t_{k}, k=1,\cdots,n),\\
  V_{B}(t,X_{t})&=&H_{k(t)}(t,X_{t}).
\end{eqnarray*}
Calculation of this expected value is the most difficult issue to evaluate Bermudan swaption and various methods have been developed.

\subsection{Fast Gauss Transform}
\label{section:FGT}
Fast Gauss transform (FGT) is a method used to calculate discrete convolution products for Gaussian functions.
\begin{eqnarray}
  g_{i}=\sum_{j=1}^{N}q_{j}\exp\left\{-\frac{\left(x_{i}-y_{j}\right)^{2}}{\delta}\right\} (i=1,2,\cdots,N'),\label{formula:def_gauss_transform}
\end{eqnarray}
It can be performed by calculating order $O(N+N')$ where direct calculation requires $O(NN')$.

\begin{figure}
  \centering
  \includegraphics[keepaspectratio, scale=0.3]{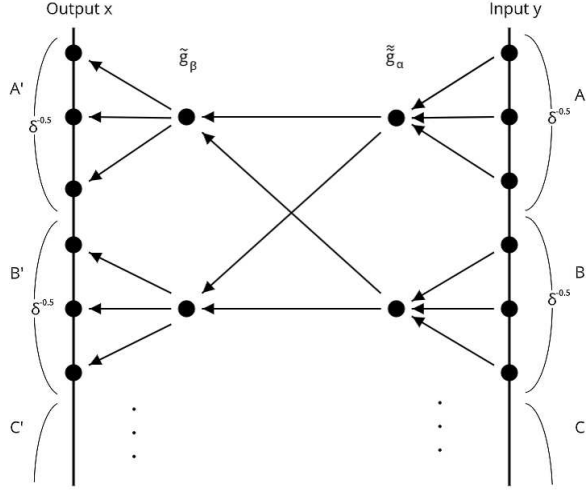}
  \caption{Fast Gauss Transform DataFlow}
  \label{figure:FGTBlock1D}
\end{figure}

Describe the principle of fast Gauss transform.
The Hermitian polynomial and its generating function are defined as follows.
\begin{eqnarray*}
  H_{\alpha}(x)&=&(-1)^{\alpha}e^{x^{2}}\frac{d^{\alpha}}{dx^{\alpha}}e^{-x^{2}},\\
  e^{-y^{2}+2xy}&=&\sum_{\alpha=0}^{\infty}H_{\alpha}(x)\frac{y^{\alpha}}{\alpha!},
\end{eqnarray*}
By multiplying both sides of the generating function by $e^{-x^{2}}$ and transforming the expression as $h_{\alpha}(x)=e^{-x^{2}}H_{\alpha}(x)$, we obtain
\begin{eqnarray*}
  e^{-(x-y)^{2}}&=&\sum_{\alpha=0}^{\infty}h_{\alpha}(x)\frac{y^{\alpha}}{\alpha!}\nonumber\\
  &=&\sum_{\alpha=0}^{\infty}h_{\alpha}(x-y_{0})\frac{(y-y_{0})^{\alpha}}{\alpha!}\nonumber\\
  &=&\sum_{\alpha=0}^{\infty}\sum_{\beta=0}^{\infty}\frac{(x_{0}-x)^{\beta}}{\beta!}h_{\alpha+\beta}(x_{0}-y_{0})\frac{(y-y_{0})^{\alpha}}{\alpha!}.
\end{eqnarray*}
The transformation of the expression from the second line to the third line is given by a Taylor expansion around $x=x_{0}$ and
\begin{eqnarray*}
  \frac{d^{\beta}}{dx^{\beta}}h_{\alpha}(x)&=&(-1)^{\alpha}\frac{d^{\alpha+\beta}}{dx^{\alpha+\beta}}e^{-x^{2}}=(-1)^{\beta}h_{\alpha+\beta}(x).
\end{eqnarray*}
We scale it by $\sqrt{\delta}$, we obtain
\begin{eqnarray*}
  e^{-\frac{(x-y)^{2}}{\delta}}&=&\sum_{\alpha=0}^{\infty}\sum_{\beta=0}^{\infty}\frac{1}{\alpha!}\frac{1}{\beta!}\left(\frac{x_{0}-x}{\sqrt{\delta}}\right)^{\beta}h_{\alpha+\beta}\left(\frac{x_{0}-y_{0}}{\sqrt{\delta}}\right)\left(\frac{y-y_{0}}{\sqrt{\delta}}\right)^{\alpha}.
\end{eqnarray*}
This infinite sum converges fast for the number of $\alpha,\beta$.
According to Broadie and Yamamoto \cite{broadie2003application}, we only need to sum up to $\alpha=\beta=8$ to achieve an error of $10^{-8}$ under $\left|\frac{x-x_{0}}{\sqrt{\delta}}\right|<\frac{1}{2}$,$\left|\frac{y-y_{0}}{\sqrt{\delta}}\right|<\frac{1}{2}$.
However, we will use different settings in numerical experiment for some reason.
Using this expansion, the convolution (\ref{formula:def_gauss_transform}) is written by
\begin{eqnarray}
  g_{i}&=&\sum_{j=1}^{N}q_{j}\exp\left\{-\frac{\left(x_{i}-y_{j}\right)^{2}}{\delta}\right\}\nonumber\\
  &=&\sum_{j=1}^{N}q_{j}\sum_{\alpha=0}^{\infty}\sum_{\beta=0}^{\infty}\frac{1}{\alpha!}\frac{1}{\beta!}\left(\frac{x_{0}-x_{i}}{\sqrt{\delta}}\right)^{\beta}h_{\alpha+\beta}\left(\frac{x_{0}-y_{0}}{\sqrt{\delta}}\right)\left(\frac{y_{j}-y_{0}}{\sqrt{\delta}}\right)^{\alpha}\nonumber\\
  &=&\sum_{\beta=0}^{\infty}\frac{1}{\beta!}\left(\frac{x_{0}-x_{i}}{\sqrt{\delta}}\right)^{\beta}\tilde{g}_{\beta},\\
  \tilde{g}_{\beta}&=&\sum_{\alpha=0}^{\infty}\frac{1}{\alpha!}h_{\alpha+\beta}\left(\frac{x_{0}-y_{0}}{\sqrt{\delta}}\right)\tilde{\tilde{g}}_{\alpha},\label{formula:FGTTrans}\\
  \tilde{\tilde{g}}_{\alpha}&=&\sum_{j=1}^{N}q_{j}\left(\frac{y_{j}-y_{0}}{\sqrt{\delta}}\right)^{\alpha},
\end{eqnarray}
here, $\sum_{j=1}^{N}q_{j}\left(\frac{y_{j}-y_{0}}{\sqrt{\delta}}\right)^{\alpha}$ does not depend on $x_{i}$ and $\beta$,
so it can be performed by calculating order $O(N+N')$ with saving the result of the calculation in advance.
When not all the points are close enough to $x_{0},y_{0}$,
we can divide the points into appropriate blocks as shown in Figure \ref{figure:FGTBlock1D}.
Let $A,B,C,\cdots$ be blocks of inputs and $A',B',C',\cdots$ be blocks of outputs.
In addition, let $y_{j}^{A},y_{j}^{B},y_{j}^{C}\cdots$ and $x_{j}^{A'},x_{j}^{B'},x_{j}^{C'}\cdots$ for $j=1\cdots N$ are points included in input and output blocks. 
Then, we can select $x_{0},y_{0}$ in each block and blocked FGT is calculated by the following three steps.
Firstly, we calculate $\tilde{\tilde{g}}_{\alpha}$ in each input block as
\begin{eqnarray}
  \tilde{\tilde{g}}_{\alpha}^{X} &=& \sum_{j=1}^{N}q_{j}\left(\frac{y_{j}^X-y_{0}^X}{\sqrt{\delta}}\right)^{\alpha},
\end{eqnarray}
where $X = A, B, C, \cdots$. Next, we calculate $\tilde{g}_{\beta}$ in each output block as
\begin{eqnarray}
  \tilde{g}_{\beta}^{X'} &=& \sum_{\alpha=0}^{\infty} \frac{1}{\alpha!} h_{\alpha+\beta}\left(\frac{x_{0}^{X'}-y_{0}^{A}}{\sqrt{\delta}}\right)\tilde{\tilde{g}}_{\alpha}^A
  + \sum_{\alpha=0}^{\infty} \frac{1}{\alpha!} h_{\alpha+\beta}\left(\frac{x_{0}^{X'}-y_{0}^{B}}{\sqrt{\delta}}\right)\tilde{\tilde{g}}_{\alpha}^B+\cdots,
\end{eqnarray}
where $X'=A',B',C',\cdots$. Lastly, we have
\begin{eqnarray}
  g_{i}^{X'}&=&\sum_{\beta=0}^{\infty}\frac{1}{\beta!}\left(\frac{x_{0}^{X'}-x_{i}^{X'}}{\sqrt{\delta}}\right)^{\beta}\tilde{g}_{\beta}^{X'}.
\end{eqnarray}

In the case of using block partitioning,
a direct calculation requires $x$ block size times $y$ block size of (\ref{formula:FGTTrans}).
However, considering that the Gaussian function decreases rapidly and we can ignore if $|x_{0}-y_{0}|$ is sufficiently large, for example larger than $8\sqrt{\delta}$,
we can calculate by the same computational complexity $O(N+N')$ when dividing into blocks.

Although the FGT is superior to the direct computation in terms of computational complexity,
the coefficient of computational complexity is large.
It may be better to use the direct computation when there are only a few points in the block.

The two-dimensional case can also be derived using the product of one-dimensional expressions
\begin{eqnarray*}
  g_{2,i}&=&\sum_{j=1}^{N}q_{2,j}\exp\left\{-\frac{\left(x_{0,i}-y_{0,j}\right)^{2}}{\delta}\right\}\exp\left\{-\frac{\left(x_{1,i}-y_{1,j}\right)^{2}}{\delta}\right\}\\
  &=&\sum_{\beta_{0}=0}^{\infty}\sum_{\beta_{1}=0}^{\infty}\frac{1}{\beta_{0}!}\frac{1}{\beta_{1}!}\left(\frac{x_{0,0}-x_{0,i}}{\sqrt{\delta}}\right)^{\beta_{0}}\left(\frac{x_{1,0}-x_{1,i}}{\sqrt{\delta}}\right)^{\beta_{1}}\tilde{g}_{2,\beta_{0},\beta_{1}},\\
  \tilde{g}_{2,\beta_{0},\beta_{1}}&=&\sum_{\alpha_{0}=0}^{\infty}\sum_{\alpha_{1}=0}^{\infty}\frac{1}{\alpha_{0}!}\frac{1}{\alpha_{1}!}h_{\alpha_{0}+\beta_{0}}\left(\frac{x_{0,0}-y_{0,0}}{\sqrt{\delta}}\right)h_{\alpha_{1}+\beta_{1}}\left(\frac{x_{1,0}-y_{1,0}}{\sqrt{\delta}}\right)\tilde{\tilde{g}}_{2,\alpha_{0},\alpha_{1}},\\
  \tilde{\tilde{g}}_{2,\alpha_{0},\alpha_{1}}&=&\sum_{j=1}^{N}q_{2,j}\frac{1}{\alpha_{0}!]}\frac{1}{\alpha_{1}!}\left(\frac{y_{0,j}-y_{0,0}}{\sqrt{\delta}}\right)^{\alpha_{0}}\left(\frac{y_{1,j}-y_{1,0}}{\sqrt{\delta}}\right)^{\alpha_{1}},
\end{eqnarray*}
where $(i=1,2,\cdots,N')$. This can be performed with the same computational complexity $O(N+N')$.
In the 2D case, as in the 1D case, we divide $x$ and $y$ into appropriate regions and apply the FGT to the calculations between each block.
We can similarly extend to more than three dimensions.
As increasing dimension number $d$, intermediate variables like $\tilde{\tilde{g}}_{2,\alpha_{0},\alpha_{1}}, \tilde{g}_{2,\beta_{0},\beta_{1}}$ increase exponential power of $d$.
For more than three dimensions, improved method is proposed by Yang et al. \cite{yang2003improved}.

\section{Pricing Bermudan Swaption using Fast Gauss Transform}

\subsection{Discretization of Expectation with FGT}
In this subsection, we show that the expected value calculation that appears in the recurrence relation of the Bermudan swaption can be discretized by numerical integration in the two factor Hull-White model,
and that the FGT can be applied to the discretization to reduce calculation time.
In addition, we propose a method of grid rotation for numerical integration
in order to reduce the discretization error and increase the efficiency of the FGT,

For the state variable $\bm{X}_{t}$ of the two factor Hull-White model, we introduce the $t$-dependent coordinate rotation as
\begin{eqnarray*}
  \bm{Y}_{t}&=&
  \begin{pmatrix}
    Y_{0,t}\\
    Y_{1,t}
  \end{pmatrix}
  =\bm{\Xi}(t)\bm{X}_{t},\\
  \bm{\Xi}(t)&=&
  \begin{pmatrix}
    \cos\left(\xi(t)\right) & -\sin\left(\xi(t)\right)\\
    \sin\left(\xi(t)\right) & \cos\left(\xi(t)\right)
  \end{pmatrix}.
\end{eqnarray*}

The relationship between $\bm{Y}_{t}$ and $\bm{Y}_{\tau}$ is deduced from (\ref{formula:ForwardX}) as
\begin{eqnarray*}
  \bm{Y}_{\tau}&=&\hat{\bm{\mu}}(t,\tau)\bm{Y}_{t}+\hat{\bm{\eta}}(t,\tau)+\hat{\bm{U}}^{\tau}(t,\tau),\\
  \hat{\bm{\mu}}(t,\tau)&=&\bm{\Xi}(\tau)\bm{\mu}(t,\tau)\bm{\Xi}^{T}(t),\\
  \hat{\bm{\eta}}(t,\tau)&=&\bm{\Xi}(\tau)\bm{\eta}(t,\tau),\\
  \hat{\bm{U}}^{\tau}(t,\tau)&=&\bm{\Xi}(\tau)\bm{U}^{\tau}(t,\tau).
\end{eqnarray*}

The variance-covariance matrix of $\hat{\bm{U}}^{\tau}(t,\tau)$ is
\begin{eqnarray}
  \hat{\bm{\Sigma}}(t,\tau)&=&\bm{\Xi}(\tau)\bm{\Sigma}(t,\tau)\bm{\Xi}^{T}(\tau), \label{formula:CovTrans}
\end{eqnarray}
and let $\hat{\bm{C}}(t,\tau)$ be its Cholesky decomposition as
\begin{eqnarray*}
  \hat{\bm{\Sigma}}(t,\tau)=\hat{\bm{C}}(t,\tau)\hat{\bm{C}}^{T}(t,\tau).
\end{eqnarray*}
Since the probability distribution of $\hat{\bm{U}}^{\tau}(t,\tau)$ under $\mathcal{F}_{t}$ follows a multivariate normal distribution,
the transition density function from $\bm{Y}_{t}=\bm{y}_{t}=(y_{0,t}, y_{1,t})^{T}$ to $\bm{Y}_{\tau}=\bm{y}_{\tau}=(y_{0,\tau}, y_{1,\tau})^{T}$ is expressed by
\begin{eqnarray*}
  \phi(\bm{y}_{\tau};\bm{y}_{t})&=&
  \frac{1}{2\pi|\hat{\bm{\Sigma}}(t,\tau)|^{\frac{1}{2}}}\exp\left\{-\frac{1}{2}\bm{z}^{T}\hat{\bm{\Sigma}}^{-1}(t,\tau)\bm{z}\right\},\\
  \bm{z}&=&\bm{y}_{\tau}-\hat{\mu}(t,\tau)\bm{y}_{t}-\hat{\eta}(t,\tau).
\end{eqnarray*}

Renaming $\hat{V}(t,\bm{Y}_{t})=V(t,\bm{X}_{t})$, (\ref{formula:G2Evaluation}) can be transformed as
\begin{eqnarray*}
  \hat{V}(t,\bm{Y}_{t})&=&F(t,\tau)e^{-\bm{1}^{T}\bm{\nu}(t,\tau)\bm{\Xi}^{T}(t)\bm{Y}_{t}}\mathbb{E}_{t}^{\tau}\left[\hat{V}(\tau,\bm{Y}_{\tau})\right],\\
  \mathbb{E}_{t}^{\tau}\left[\hat{V}(\tau,\bm{Y}_{\tau})\right]&=&
  \int_{-\infty}^{\infty}\int_{-\infty}^{\infty}\hat{V}(\tau,\bm{y}_{\tau})\phi(\bm{y}_{\tau};\bm{Y}_{t})dy_{0,\tau}dy_{1,\tau}\\
  &\simeq&\sum_{i=1}^{N_{\tau}}w_{i}\hat{V}(\tau,\bm{y}_{\tau,i})\phi(\bm{y}_{\tau,i};\bm{Y}_{t}),
\end{eqnarray*}
where derivative value is approximately computed by numerical integration with $\bm{y}_{\tau,i}$ and $w_{i}$ as nodes and weights.
Furthermore, when we calculate the prices for various states $\bm{Y}_{t}=\bm{y}_{t,j}$, $(j=1,\cdots,N_{t})$ at time $t$,
we use the same nodes and weights $\bm{y}_{\tau,i},w_{i}$ for the numerical integration. Therefore, we obtain the approximation as
\begin{eqnarray}
  \hat{V}(t,\bm{y}_{t,j})&\simeq&F(t,\tau)e^{-\bm{1}^{T}\bm{\nu}(t,\tau)\bm{\Xi}^{T}(t)\bm{y}_{t,j}}\sum_{i=1}^{N_{\tau}}w_{i}\hat{V}(\tau,\bm{y}_{\tau,i})\phi(\bm{y}_{\tau,i};\bm{y}_{t,j}).\label{formula:G2FGT}
\end{eqnarray}

To apply the FGT to this approximation, we introduce new variables
\begin{eqnarray*}
  \bm{n}_{\tau,i}&=&
  \begin{pmatrix}
    n_{0,\tau,i}\\
    n_{1,\tau,i}
  \end{pmatrix}=
  \hat{\bm{C}}^{-1}(t,\tau)\bm{y}_{\tau,i},\\
  \bm{m}_{t,j}&=&
  \begin{pmatrix}
    m_{0,t,j}\\
    m_{1,t,j}
  \end{pmatrix}=
  \hat{\bm{C}}^{-1}(t,\tau)
  \left\{
  \hat{\bm{\mu}}(t,\tau)\bm{y}_{t,j}
  +\hat{\bm{\eta}}(t,\tau)
  \right\},
\end{eqnarray*}
and the equation can be written as follows
\begin{eqnarray*}
  \phi(\bm{y}_{\tau,i};\bm{y}_{t,j})&=&\frac{1}{2\pi|\hat{\bm{\Sigma}}(t,\tau)|^{\frac{1}{2}}}\exp\left\{-\frac{(n_{0,\tau,i}-m_{0,t,j})^{2}+(n_{1,\tau,i}-m_{1,t,j})^{2}}{2}\right\},
\end{eqnarray*}
where $|\cdot|$ is a determinant of a matrix.
As a result, the equation (\ref{formula:G2FGT}) can be computed quickly using a two-dimensional FGT.

We divide $(n_{0,\tau,i}, n_{1,\tau,i}),(m_{0,\tau,i}, m_{1,\tau,i})$ into square blocks of width $\sqrt{2}$ and apply the FGT for each block combination
if the distance between the center points of the block is less than $\sigma_{max}$, otherwise we omit the calculation as the contribution from far block is sufficiently small.

\subsection{Numerical Integration Settings and Grid Rotation}

As an example of a specific numerical integration, we propose the following setting.
The random variable $\bm{Y}_{\tau}$ is distributed around the origin $(0,0)$ where the variance is $\hat{\bm{\Sigma}}_{00}(0,\tau),\hat{\bm{\Sigma}}_{11}(0,\tau)$ under the risk-neutral measure $\mathbb{Q}$.

For a sufficiently large $M$ and the number of partitions $N_{y}$, we set the region of
\begin{eqnarray*}
A_{+}=\left[-M_{+}\sqrt{\hat{\bm{\Sigma}}_{00}(0,\tau)},M_{+}\sqrt{\hat{\bm{\Sigma}}_{00}(0,\tau)}\right]\times
\left[-M_{+}\sqrt{\hat{\bm{\Sigma}}_{11}(0,\tau)},M_{+}\sqrt{\hat{\bm{\Sigma}}_{11}(0,\tau)}\right],
\end{eqnarray*}
where $M_{+}=\frac{M(N_{y}+1)}{N_{y}}$, in the $y_{0,\tau}$ and $y_{1,\tau}$ directions respectively,
and apply the midpoint rule of numerical integration.

In other words, we set rectangular area
\begin{eqnarray*}
  A_{-}=\left[-M\sqrt{\hat{\bm{\Sigma}}_{00}(0,\tau)},M\sqrt{\hat{\bm{\Sigma}}_{00}(0,\tau)}\right]\times
  \left[-M\sqrt{\hat{\bm{\Sigma}}_{11}(0,\tau)},M\sqrt{\hat{\bm{\Sigma}}_{11}(0,\tau)}\right],
\end{eqnarray*}
and divide this area into $N_{\tau}$ square blocks whose width is $\Delta y_{0,\tau}=\frac{2M}{N_{y}}\sqrt{\hat{\bm{\Sigma}}_{00}(0,\tau)}$,$\Delta y_{1,\tau}=\frac{2M}{N_{y}}\sqrt{\hat{\bm{\Sigma}}_{11}(0,\tau)}$.
Let $\bm{y}_{\tau,i}$, $i=1,\cdots,(N_{y}+1)^{2}$ be the nodes of numerical integration that aligns the rectangular area nodes and 
$w_{i}=\Delta y_{0,\tau}\Delta y_{1,\tau}=\frac{4M^2\sqrt{\hat{\bm{\Sigma}}_{00}(0,\tau)\hat{\bm{\Sigma}}_{11}(0,\tau)}}{N_{y}^{2}}$ be the weights.
Since the convergence rate of midpoint rule is $O(h^{2})$ where $h$ is discretization step size, 
this scheme converges $O(N_{y}^{-2})=O(N_{\tau}^{-1})$ which is faster than $O(N_{path}^{-0.5})$ of Monte-Carlo method where $N_{path}$ is the number of paths.

If we consider the grid interval in the $y_{0,\tau},y_{1,\tau}$ direction in terms of $n_{\tau}$ coordinates, we obtain following vectors
\begin{eqnarray*}
  \Delta_{y_{0,\tau}}n_{\tau}&=&
  \hat{\bm{C}}^{-1}(t,\tau)
  \begin{pmatrix}
    \Delta y_{0,\tau}\\
    0
  \end{pmatrix}=
  \frac{2M}{N_{y}}\frac{\sqrt{\hat{\bm{\Sigma}}_{00}(0,\tau)}}{\sqrt{\hat{\bm{\Sigma}}_{00}(t,\tau)}}
  \begin{pmatrix}
    1\\
    -\frac{\hat{\bm{\Sigma}}_{10}(t,\tau)}{\sqrt{\left|\hat{\Sigma}(t,\tau)\right|}}
  \end{pmatrix},\\
  \Delta_{y_{1,\tau}}n_{\tau}&=&
  \hat{\bm{C}}^{-1}(t,\tau)
  \begin{pmatrix}
    0\\
    \Delta y_{1,\tau}
  \end{pmatrix}=
  \frac{2M}{N_{y}}
  \frac{\sqrt{\hat{\bm{\Sigma}}_{00}(t,\tau)}}{\sqrt{\hat{\bm{\Sigma}}_{00}(0,\tau)}}
  \frac{\sqrt{\left|\hat{\bm{\Sigma}}(0,\tau)\right|+\hat{\bm{\Sigma}}_{10}^{2}(0,\tau)}}{\sqrt{\left|\hat{\Sigma}(t,\tau)\right|}}
  \begin{pmatrix}
    0\\
    1
  \end{pmatrix}.
\end{eqnarray*}
By approximating $\hat{\Sigma}(t,\tau)\simeq \lambda \hat{\Sigma}(0,\tau)$, we get
\begin{eqnarray*}
  \Delta_{y_{0,\tau}}n_{\tau}&\simeq&
  \frac{2M}{N_{y}\sqrt{\lambda}}
  \begin{pmatrix}
    1\\
    -\frac{\hat{\bm{\Sigma}}_{10}(t,\tau)}{\sqrt{\left|\hat{\Sigma}(t,\tau)\right|}}
  \end{pmatrix},\\
  \Delta_{y_{1,\tau}}n_{\tau}&\simeq&
  \frac{2M\sqrt{\lambda}}{N_{y}}\sqrt{1+\frac{\hat{\bm{\Sigma}}_{10}^{2}(0,\tau)}{\left|\hat{\bm{\Sigma}}(0,\tau)\right|}}
  \begin{pmatrix}
    0\\
    1
  \end{pmatrix}.
\end{eqnarray*}
Improving the accuracy of the numerical integration, we determine the rotation $\bm{\Xi}(\tau)$ so that the width between the grid points is minimized, 
which means the condition $\hat{\bm{\Sigma}}_{10}(0,\tau)=0$.
Expanding this condition, we obtain
\begin{eqnarray*}
\hat{\Sigma}_{10}(0,\tau)&=&\frac{\Sigma_{00}(0,\tau)-\Sigma_{11}(0,\tau)}{2}\sin2\xi(\tau)+\Sigma_{10}(0,\tau)\cos2\xi(\tau)=0,
\end{eqnarray*}
and value of $\xi$ as
\begin{eqnarray}
  \xi(\tau)&=&\frac{1}{2}\tan^{-1}\left(\frac{-2\Sigma_{10}(0,\tau)}{\Sigma_{00}(0,\tau)-\Sigma_{11}(0,\tau)}\right)\pm \frac{m}{2}\pi, \ \ m\in\mathbb{N},\label{formula:angle}
\end{eqnarray}
(If $\Sigma_{00}(0,\tau)=\Sigma_{11}(0,\tau)$, $\xi(\tau)=\frac{\pi}{4}\pm\frac{m}{2}\pi$ satisfies the equation.)

\begin{figure}
  \centering
  \includegraphics[width=15cm]{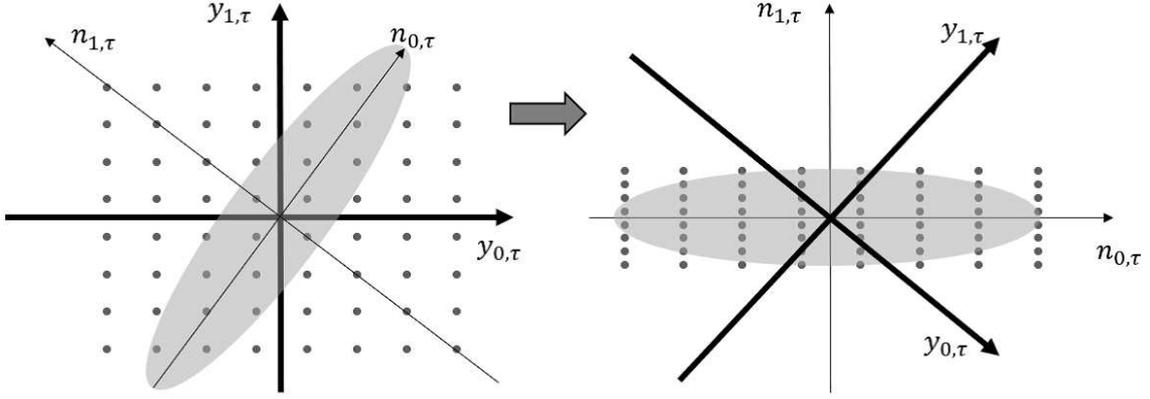}
  \caption{Grid Rotation}
  \label{figure:GridRotation}
\end{figure}

The area of the parallelogram formed by the vectors $\Delta_{y_{0,\tau}}n_{\tau},\Delta_{y_{1,\tau}}n_{\tau}$ is calculated by
\begin{eqnarray}
  \Delta_{y_{\tau}}n_{\tau}=\frac{4M^{2}}{N_{y}^{2}}
  \frac{\sqrt{\left|\hat{\bm{\Sigma}}(0,\tau)\right|+\hat{\bm{\Sigma}}_{10}^{2}(0,\tau)}}{\sqrt{\left|\hat{\Sigma}(t,\tau)\right|}},\label{formula:delta_area}
\end{eqnarray}
and it shows that $\hat{\bm{\Sigma}}_{10}(0,\tau)=0$ is also the condition that minimizes the parallelogram area.
This leads that grid points are most densely distributed as shown in Figure \ref{figure:GridRotation}.
In the calculation of the FGT, the more points contained in one block, the more efficient it becomes,
so this condition also has the effect of making the FGT more efficient.

\subsection{Applying Calculation of Expected Exposure}
CVA is a price adjustment when a counterparty defaults during the contract period.
It plays an important role in the price quotation and risk management of derivatives.

Assuming that counterparty defaults and market factors are independent, 
let $\lambda_{C}(t)$ be the default intensity,
$\mbox{LGD}$ be the "$1-$ recovery rate," and the time grid be $0=s_{0}<s_{1}<\cdots<s_{n}=T$,
$V_{t}$ be the living value of the transactions contracted with the counterparty.
We define living value as the price including only living cashflows, rights and obligations, dropping cashflows that have already been paid out until $t$
and considering the transaction changes by exercise.
For example, $V_{t}$ contains only a Bermudan swaption and let $\tau_{B}$ be the stopping time which represents the optimal exercise time of the Bermudan swaption.
In the set of the events $\left\{\omega|\tau_{B}\leq t\right\}$, $V_{t}$ expresses the price of the cashflows of exercised swap ahead of time $t$.
In the set of the complementary events, $V_{t}$ expresses the price of the Bermudan swaption which is not exercised up to $t$ yet.
CVA formula is given by
\begin{eqnarray*}
  \mbox{CVA}&=&\mbox{LGD}\int_{0}^{T}\lambda_{C}(s)e^{-\int_{0}^{s}\lambda_{C}(u)du}\mathbb{E}\left[e^{-\int_{0}^{s}R_{u}du}\max\left\{V_{s},0\right\}_{+}\right]ds,
\end{eqnarray*}
where the part of the expectation in the formula is called the expected (positive) exposure.
$V_{s_{i}}$ contains various factors such as exchange rates, interest rates, stocks, etc.,
and this expected value is calculated using the Monte Carlo method.
In other words, we generate a path of $s_{i}$ state variables using the Monte Carlo method
and take the expectation by discounted value of the portfolio corresponding to the state variables.
When a portfolio contains a Bermudan swaption, the method proposed in this paper can be combined with grid interpolation (see e.g. \cite{green2015xva}).
This is a method in which grid points are set up in the state space at time $s_{i}$, the value on the grid points is obtained by the method described in this paper as preprocess.
Then the value on the path generated by the Monte Carlo method is obtained by interpolation between the values on the grid points.

\begin{figure}
  \centering
  \begin{tabular}{c}
    \begin{minipage}[t]{1.0\hsize}
      \centering
      \includegraphics[keepaspectratio, scale=0.5]{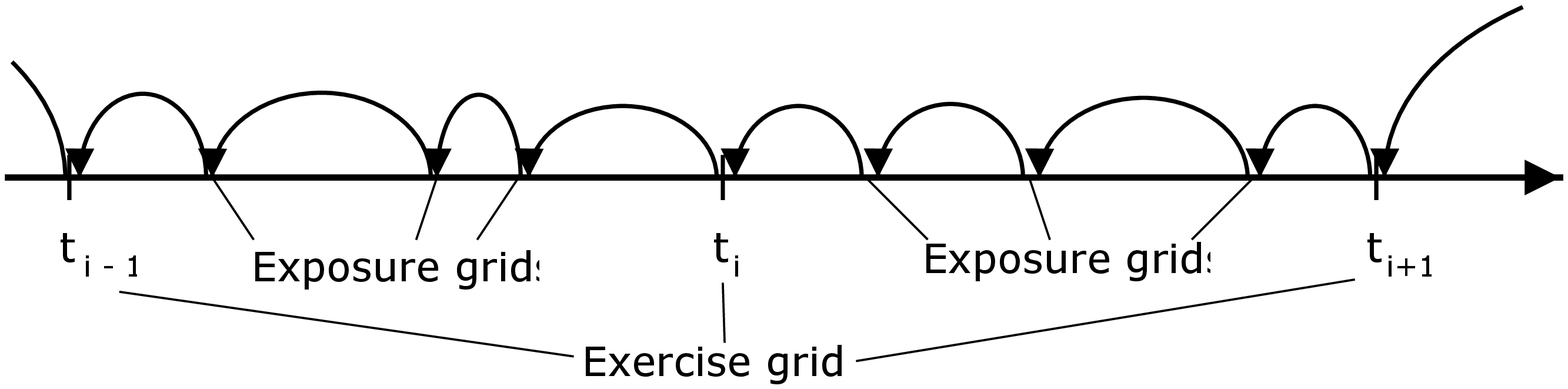}
      \caption{Step by step calculation}
      \label{figure:XVACalcStep}
    \end{minipage}\\
    \begin{minipage}[t]{1.0\hsize}
      \centering
      \includegraphics[keepaspectratio, scale=0.5]{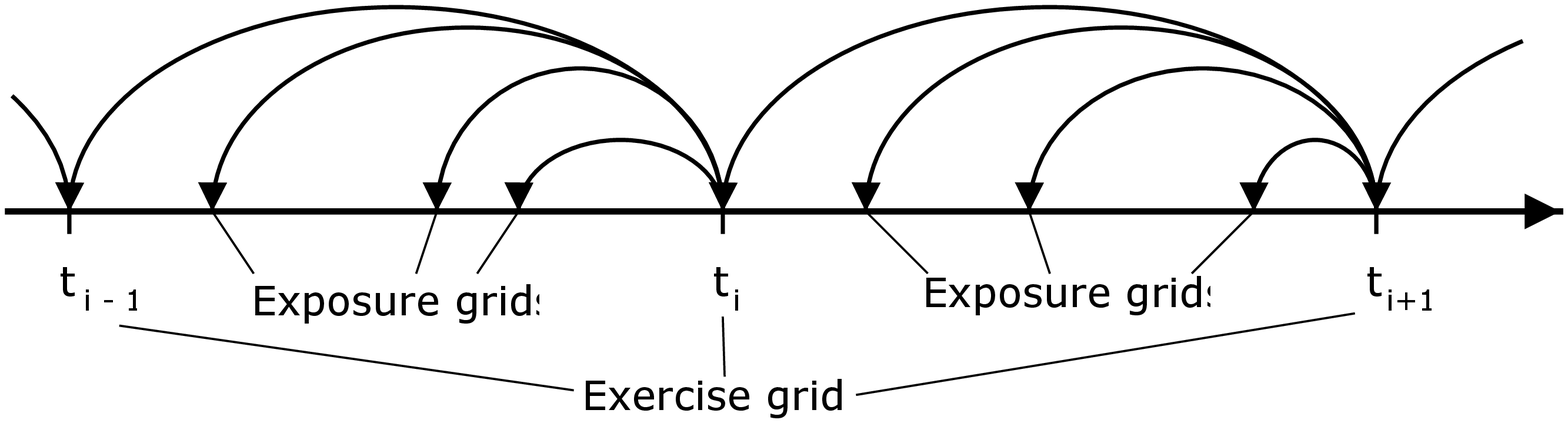}
      \caption{Leap from exercise calculation}
      \label{figure:XVACalc}
    \end{minipage}
  \end{tabular}
\end{figure}

To use this method, it is necessary to set up time grids to calculate the price at points other than the exercise.
There are two choices to calculate these points either from immediately after exposure point, as in Figure \ref{figure:XVACalcStep},
or from the next exercise point, as in Figure \ref{figure:XVACalc}.
When the time interval is narrow, the area of the equation (\ref{formula:delta_area}) becomes large,
resulting in a coarse discretization, so it is better to calculate it using Figure \ref{figure:XVACalc} method.
In addition, Figure \ref{figure:XVACalc} method has the advantage of higher parallelization efficiency
because it can calculate each expected exposures grid independently,
whereas Figure \ref{figure:XVACalcStep} method needs to be calculated sequentially.

\subsection{Extension of application}
The method proposed in this section is not limited to the computation of Bermudan swaptions by two factor Hull-White model.

Let $X_{t}$ be a state variable of arbitrary dimension,
$\mu(t,\tau),\eta(t,\tau)$ be a deterministic function,
and the same equation can be expanded by considering $U(t,\tau)$
as arbitrary multivariate normal distributions,
it can be applied to calculate the expected value $E_{t}[V(\tau,X_{\tau})]$ which payoffs can be written in terms of only the state variables at $\tau$.

A model with such a setup would apply, for example,
Gn++ model (see e.g. \cite{di2012general}) and a multi-asset model where each asset follows correlated Black-Scholes models.

The rotation matrix $\Xi$ is chosen to be an appropriate orthogonal matrix for the model.
We recommend an orthogonal matrix such that left side of formula (\ref{formula:CovTrans}) becomes close to the diagonal matrix.

\section{Numerical Example}
\subsection{Experimental Overview and Environment}
In these experiments we use two factor Hull-White model whose coefficients are constant ($\kappa_{0}(t)=\kappa_{0},\kappa_{1}(t)=\kappa_{1},\sigma_{0}(t)=\sigma_{0},\sigma_{1}(t)=\sigma_{1}$).
Other variables that appear in our method are described in Appendix \ref{sec:G2PPFormula}.
We implemented our FGT method for pricing Bermudan swaption using the C++ language
and evaluate speed and precision on Xeon Gold 6130 CPU.
The operating system was Windows Server 2016 Standard Edition, and the compiler was Microsoft Visual Studio 2017.

The conditions for calculating the Bermudan swaption price are $M=8$,
$\sigma_{max}=8$, and we choose a branch in \ref{formula:angle} as $\hat{\Sigma}_{00}(0,\tau)$ become smallest for continuity.

For FGT, the width of the block is set as $2\sqrt{\delta}$ which is twice mentioned in subsection \ref{section:FGT} to reduce
the number of inter-block calculation which become a bottleneck when the number of grid is small,
and the expansion order of the FGT is set as $\alpha_{0}=\alpha_{1}=\beta_{0}=\beta_{1}=32$ to achieve precision of double type.

Using USD market data from Bloomberg\footnote{Provided by CARF (Center for Advanced Research), The University of Tokyo} for September 20 2018 and Dec 5 2019,
we calibrated model parameters with swaptions where the sum of the maturity and the underlying term is between 4Y and 6Y.

\begin{figure}[H]
  \begin{tabular}{cc}
    \begin{minipage}[t]{0.45\hsize}
      \centering
      \includegraphics[keepaspectratio, scale=0.4]{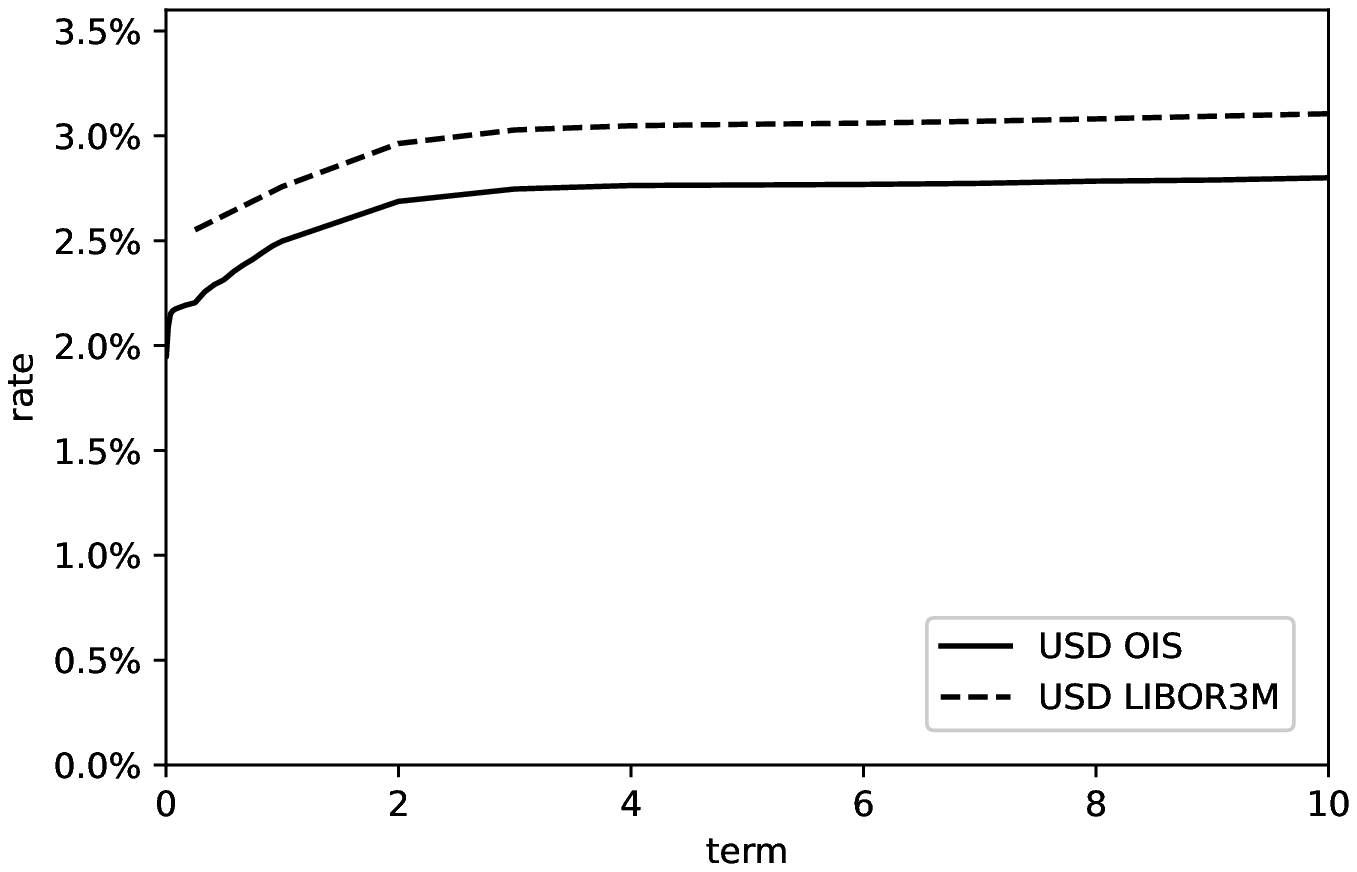}
      \caption{Sep 20 2018 USD Curve}
      \label{figure:20180920USDCurve}
    \end{minipage} &
    \begin{minipage}[t]{0.45\hsize}
      \centering
      \includegraphics[keepaspectratio, scale=0.4]{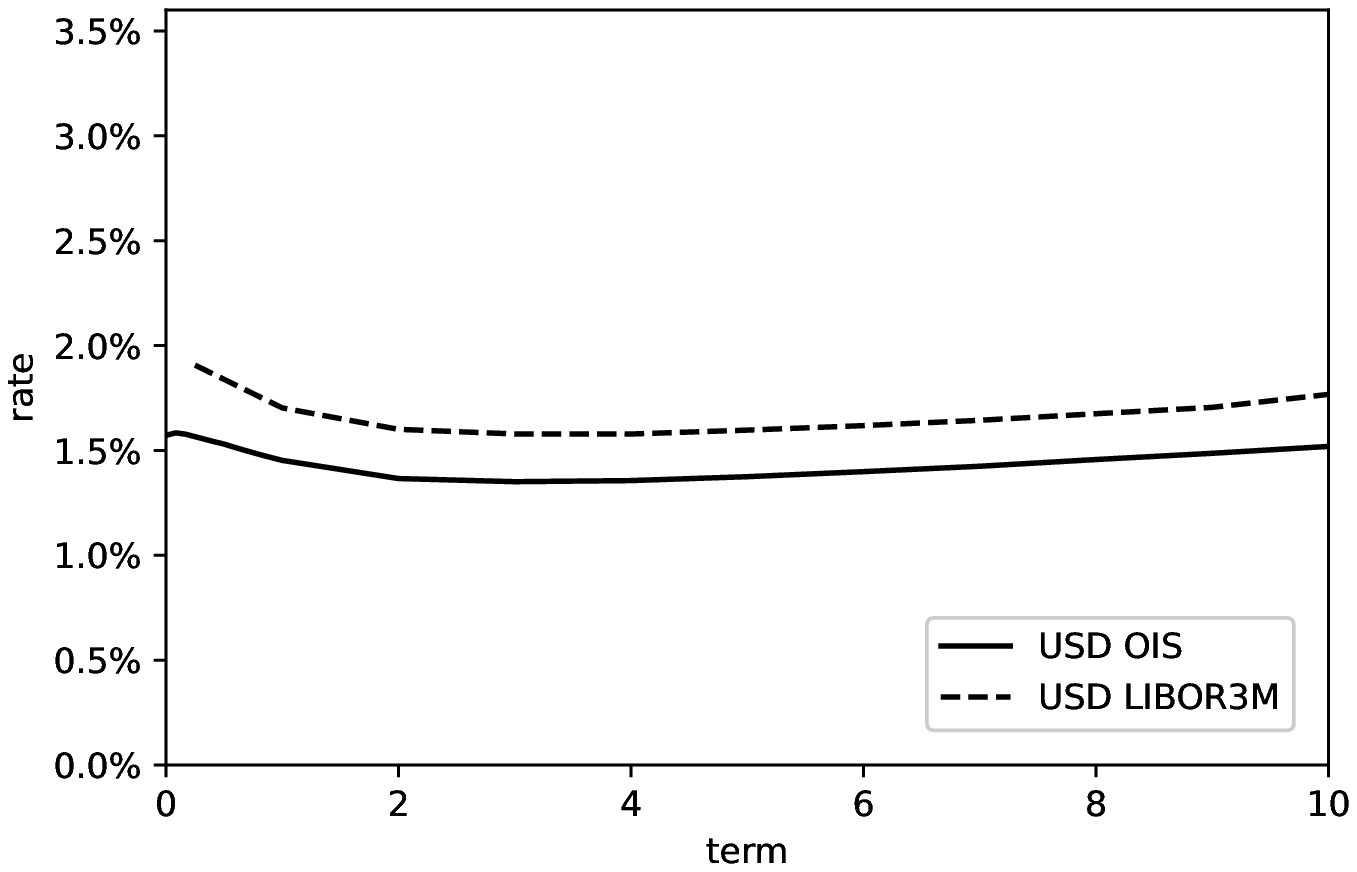}
      \caption{Dec 28 2019 USD Curve}
      \label{figure:20191205USDCurve}
    \end{minipage}
  \end{tabular}
\end{figure}

The curves of the discount rate and the LIBOR forecast rate at each reference date were in Figure \ref{figure:20180920USDCurve},\ref{figure:20191205USDCurve}.
The calibrated model parameters are in Table \ref{table:20180920G2Param},\ref{table:20191205G2Param} and these values were used in the following experiments.
Market data of September 20 2018 and December 5 2019 are chosen for the reason that correlation $\rho$ of the former is close to $-1$ and the latter is not.

\begin{table}[H]
  \centering
  \begin{tabular}{c}
    \begin{minipage}[t]{1.0\hsize}
      \centering
      \caption{Sep 2018 20 Model Parameter}
      \label{table:20180920G2Param}
      \begin{tabular}{|ccccc|}
        \hline
         $\kappa_{0}$ & $\kappa_{1}$ & $\sigma_{0}$ & $\sigma_{1}$ & $\rho$\\
        \hline
        0.764924667 & 0.352480535 & 0.064510503 & 0.043555081 & -0.988465395\\
        \hline
      \end{tabular}
    \end{minipage}\\
    \begin{minipage}[t]{1.0\hsize}
      \centering
      \caption{Dec 5 2019 Model Parameter}
      \label{table:20191205G2Param}
      \begin{tabular}{|ccccc|}
        \hline
         $\kappa_{0}$ & $\kappa_{1}$ & $\sigma_{0}$ & $\sigma_{1}$ & $\rho$\\
        \hline
        1.557180934 & 0.080090711 & 0.010574543 & 0.008692398 & -0.900422625\\
        \hline
      \end{tabular}
    \end{minipage}
  \end{tabular}
\end{table}

\subsection{Comparing the speed and accuracy of Bermudan swaption PV evaluation}
We calculated the price of a Bermudan swaption whose underlying asset is a 3-month roll for both fixed and floating legs, 
fixed rate is set near 5 years swap rate (September 20 2018 data: 3.0564\%, December 5 2019 data: 1.5875\%), floating index is USD LIBOR 3M,
and exercise date are start from 3-month forward and 3 month frequency.
To compare the efficiency, we define three settings that 
"FGT" is using FGT but no grid rotation, "FGT+Rotate" is using both FGT and grid rotation, and "NoFGT+Rotate" is only grid rotation.

The number of $N_{y}$ was prepared from 50 to 6400 (up to 600 for NoFGT+Rotate due to the calculation time)

\begin{figure}[h]
  \begin{tabular}{cc}
    \begin{minipage}[t]{0.45\hsize}
      \centering
      \includegraphics[keepaspectratio, scale=0.4]{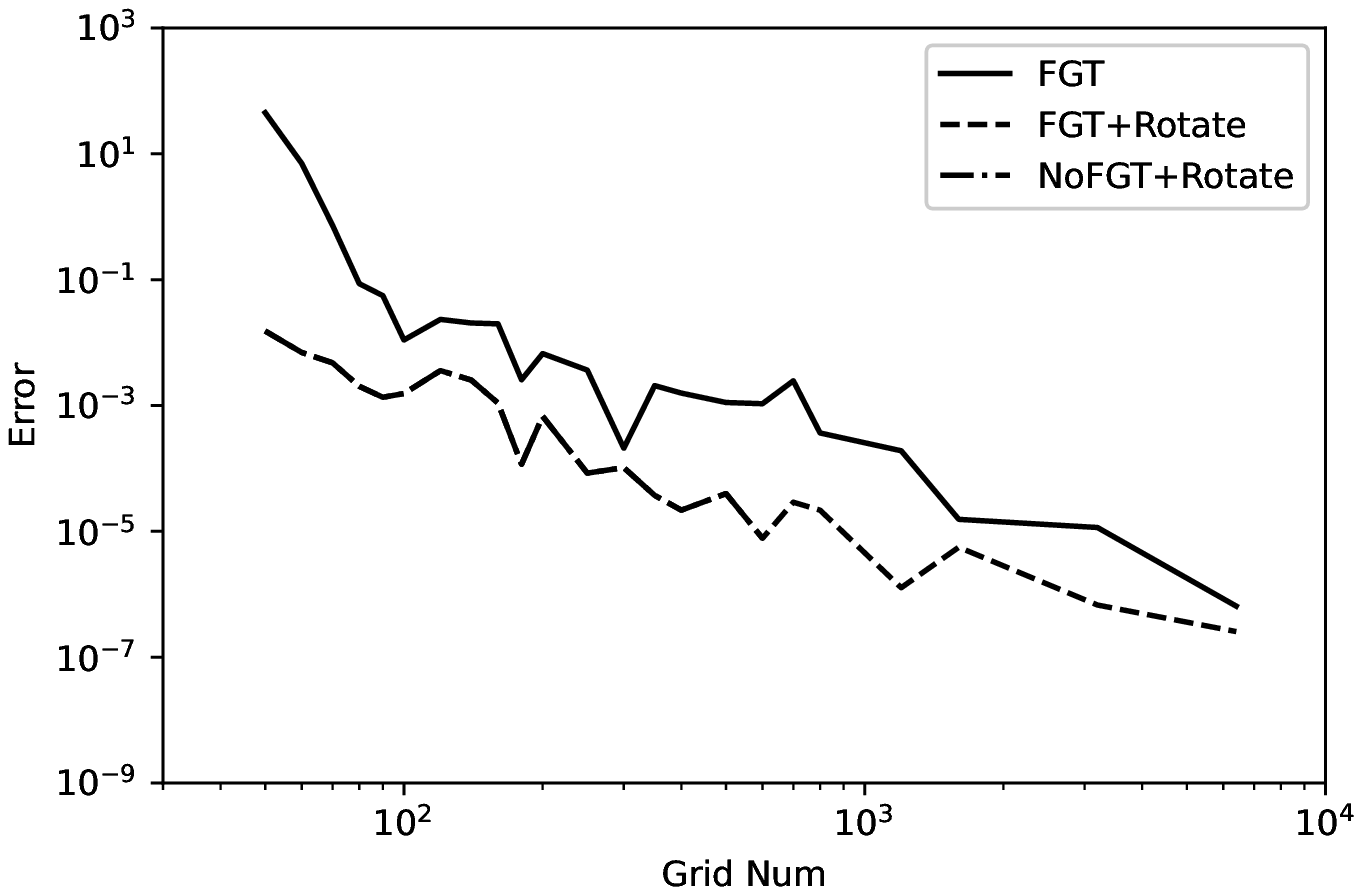}
      \caption{Sep 20 2018 Grid Num - Precision Plot}
      \label{figure:Ex20180920Grid-Precision}
    \end{minipage} &
    \begin{minipage}[t]{0.45\hsize}
      \centering
      \includegraphics[keepaspectratio, scale=0.4]{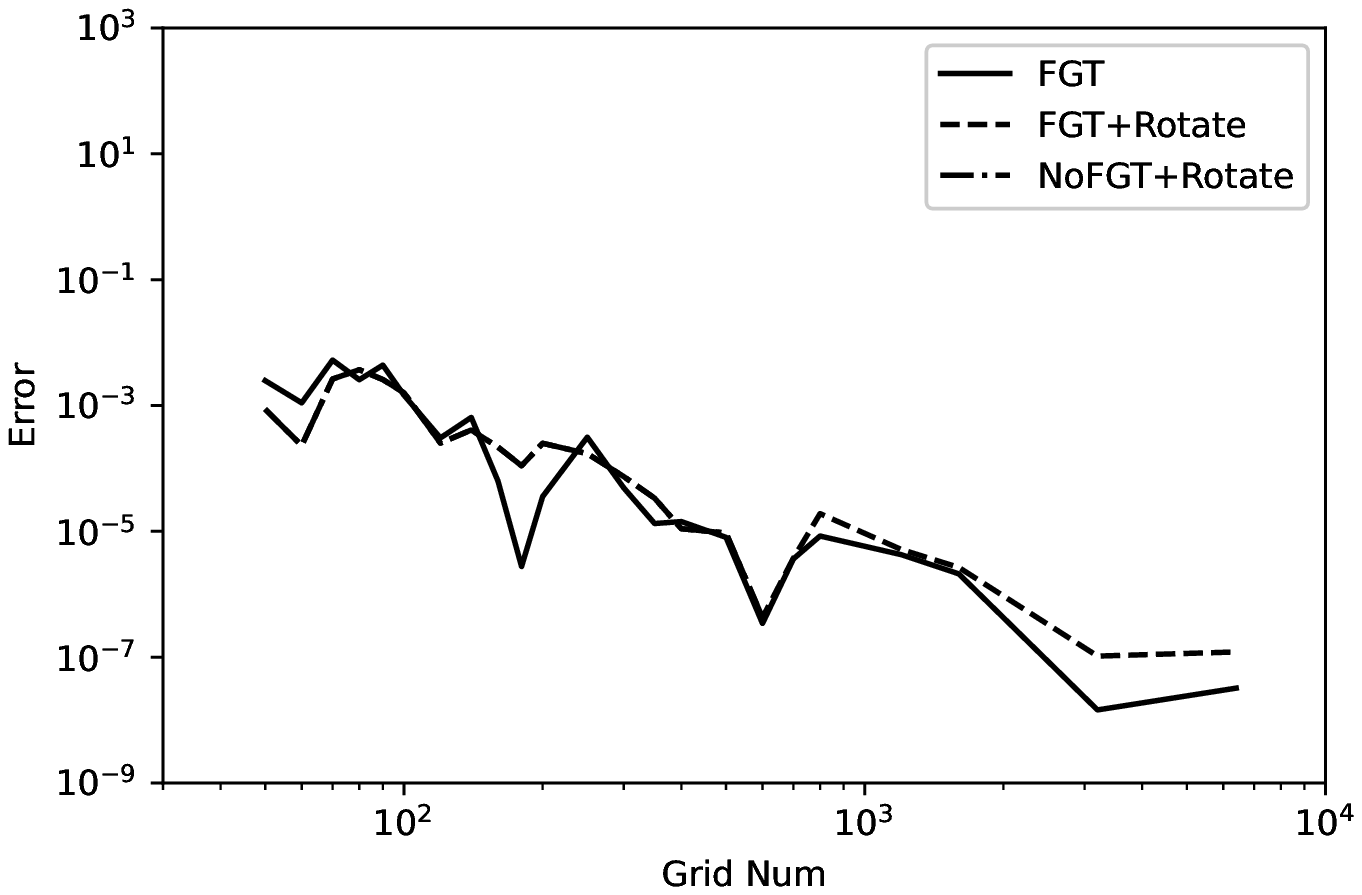}
      \caption{Dec 5 2019 Grid Num - Precision Plot}
      \label{figure:Ex20191205Grid-Precision}
    \end{minipage}
  \end{tabular}
  \begin{tabular}{cc}
    \begin{minipage}[t]{0.45\hsize}
      \centering
      \includegraphics[keepaspectratio, scale=0.4]{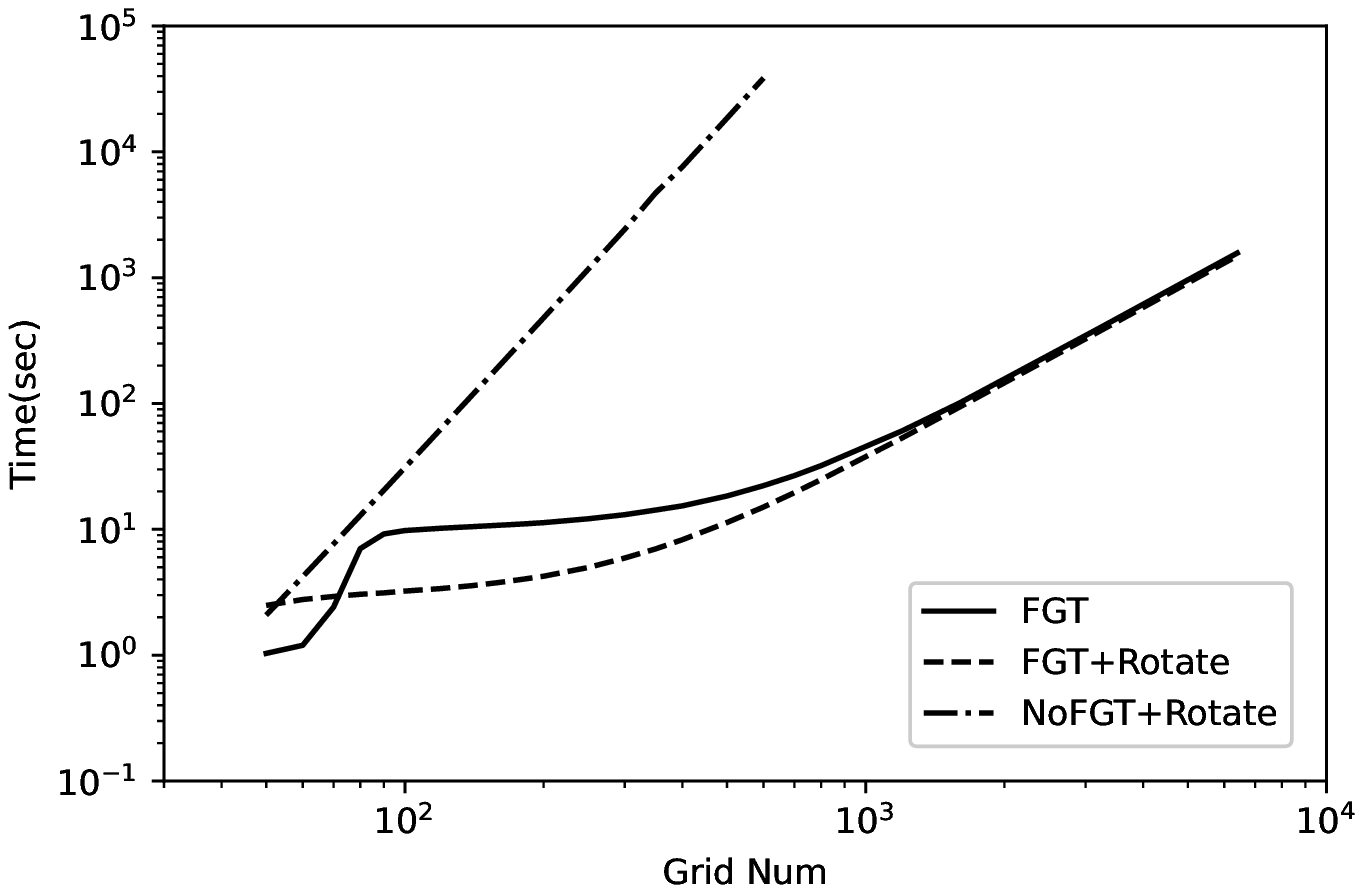}
      \caption{Sep 20 2018 Grid Num - Time Plot}
      \label{figure:Ex20180920Grid-Time}
    \end{minipage} &
    \begin{minipage}[t]{0.45\hsize}
      \centering
      \includegraphics[keepaspectratio, scale=0.4]{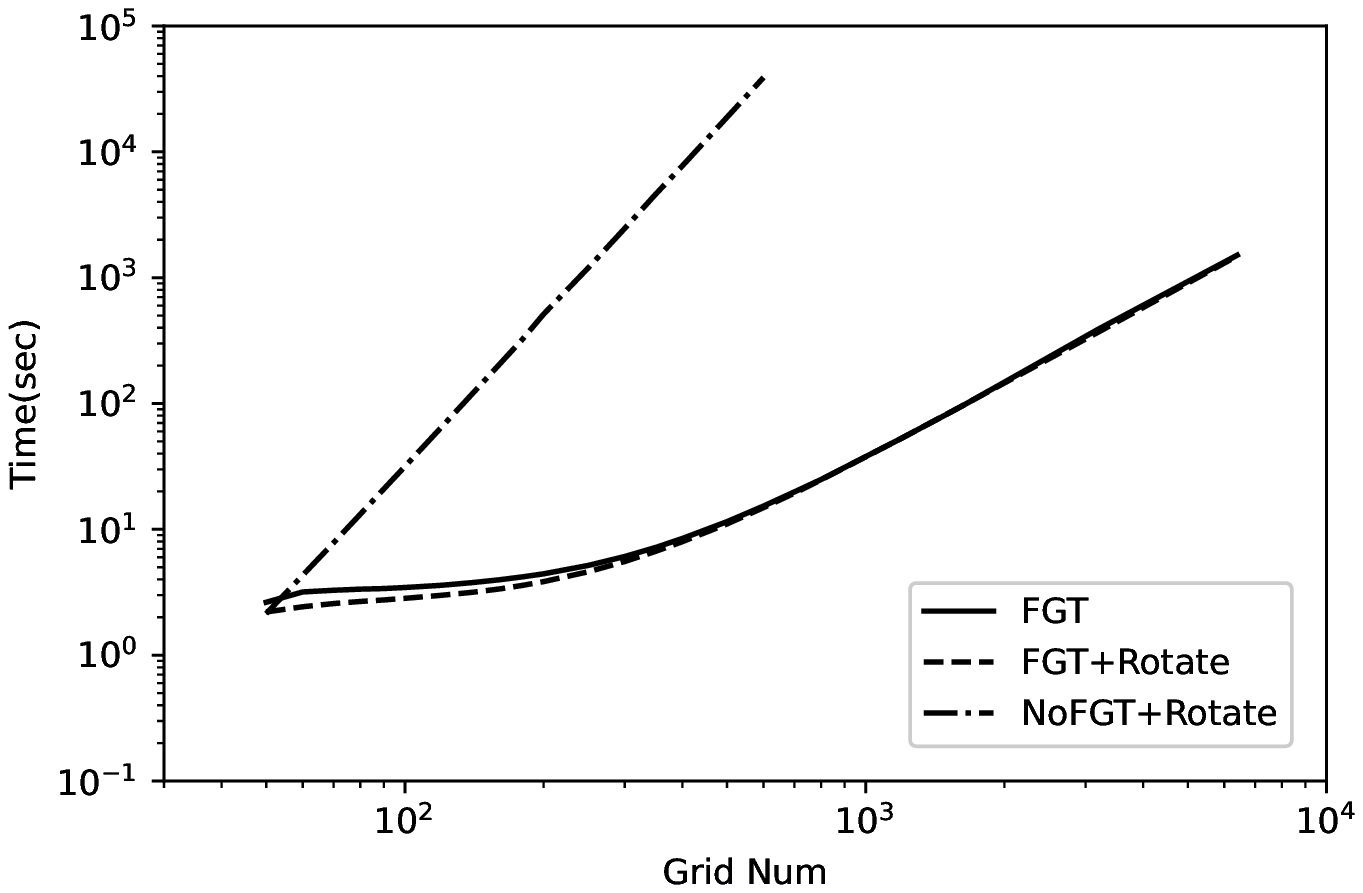}
      \caption{Dec 5 2019 Grid Num - Time Plot}
      \label{figure:Ex20191205Grid-Time}
    \end{minipage}
  \end{tabular}
  \begin{tabular}{cc}
    \begin{minipage}[t]{0.45\hsize}
      \centering
      \includegraphics[keepaspectratio, scale=0.4]{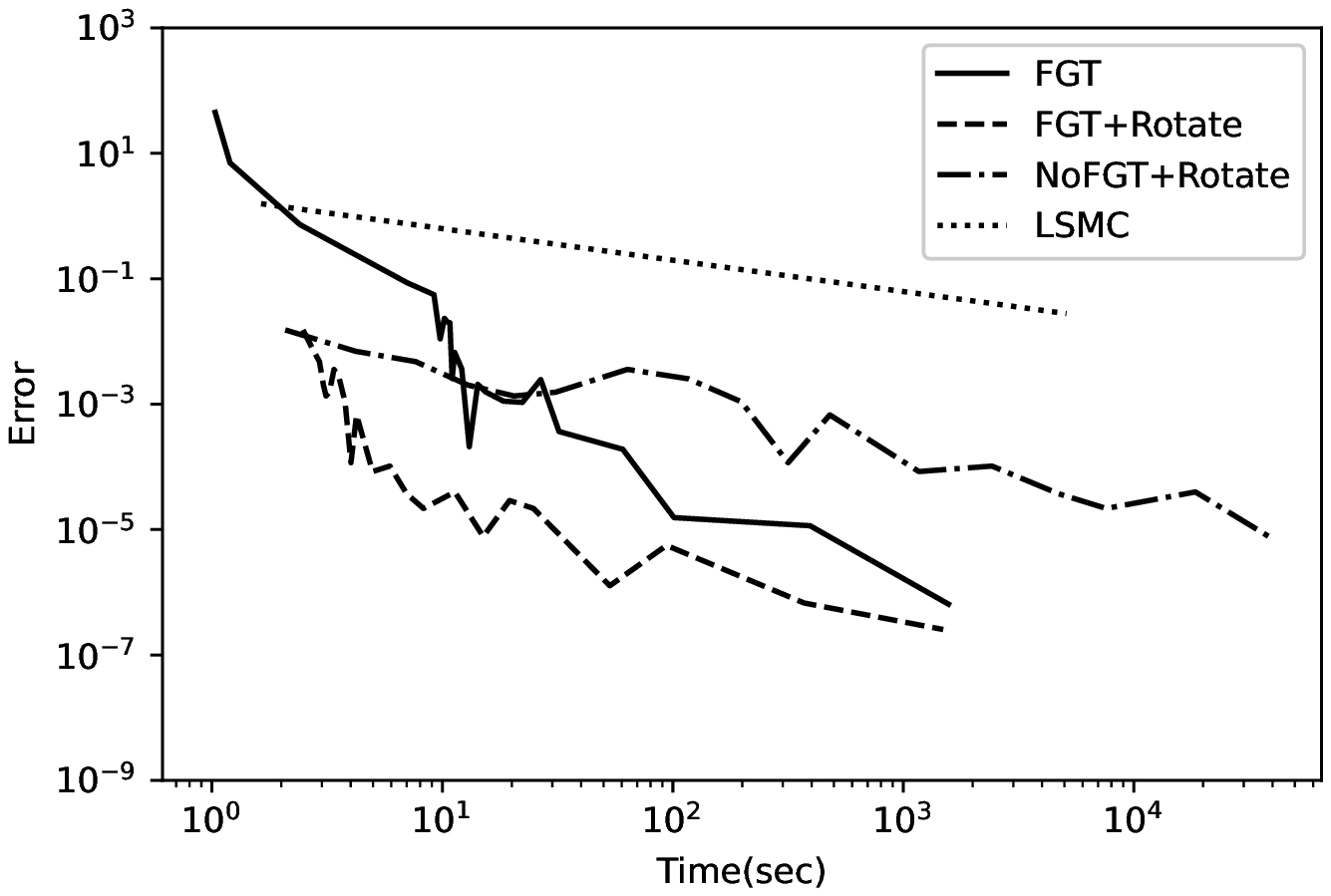}
      \caption{Sep 20 2018 Time - Precision Plot}
      \label{figure:Ex20180920Time-Precision}
    \end{minipage} &
    \begin{minipage}[t]{0.45\hsize}
      \centering
      \includegraphics[keepaspectratio, scale=0.4]{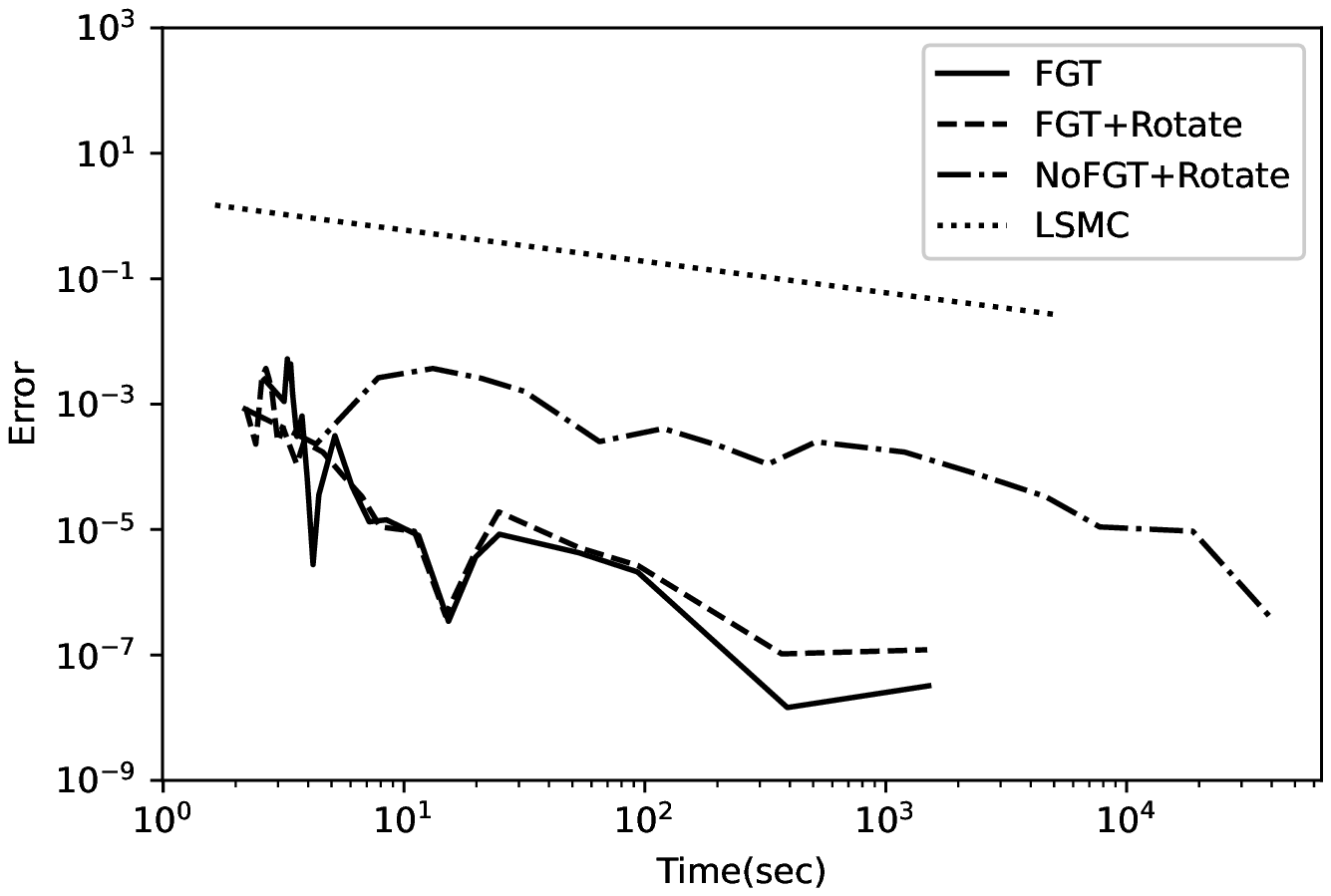}
      \caption{Dec 5 2019 Time - Precision Plot}
      \label{figure:Ex20191205Time-Precision}
    \end{minipage}
  \end{tabular}
\end{figure}

We assume the value calculated by "FGT+Rotate" with $N_{y}=25600$ as the true value
and we plot $N_{y}$ and the difference from true value in Figure \ref{figure:Ex20180920Grid-Precision}, \ref{figure:Ex20180920Grid-Precision} as log-log graph.
In these figure, "FGT+Rotate" and "NoFGT+Rotate" are in almost perfect agree
since we set $\alpha_{max},\beta_{max}$ of FGT more than the accuracy of double type.
From another perspective, Figure \ref{figure:Ex20180920Grid-Precision}, using September 20 2018 data, shows that the computational accuracy of "FGT" suddenly become worse as the grid becomes coarser,
while error of "FGT+Rotate" is suppressed even with a coarse grid.
This phenomenon is not seen in figure \ref{figure:Ex20191205Grid-Precision} using December 5 2019 data.
It is indicating that grid rotation improves stability when the grid is coarse in markets where the correlation is very close to $-1$.

Figure \ref{figure:Ex20180920Grid-Time} and Figure \ref{figure:Ex20191205Grid-Time} are log-log plots of $N_{y}$ and the computational speed for each calculation method. 
Theoretically, the computational time increases on the order of the square of $N_{y}$ when FGT is used
and on the order of the fourth power of $N_{y}$ when FGT is not used because nodes for integration is proportional to $N_{y}^{2}$.
In fact, the figure shows that the time is increasing theoretically when the number of grids is large.
In contrast, the computation time is almost no change when the grid is coarse when FGT is used.
The reason seems that the block-to-block calculations of (\ref{formula:FGTTrans}) take up a large portion of computational time
when the number of grid points in a block is small.
Therefore, an improvement to switch the use of FGT can be considered when the number of grid points in a block is below a certain level.

In addition, the result using September 20 2018 data shows a significant increase in computation time on coarse grids when grid rotation is not used.
The reason is thought that the density of grid points increases due to the increase in $\Delta\bm{n}_{\tau}$, and the number of blocks in the FGT increases
when the grid rotation is not used in the situation where the correlation is very close to $-1$.

Figure \ref{figure:Ex20180920Time-Precision} and Figure \ref{figure:Ex20191205Time-Precision} are log-log plots
with computation time on the horizontal axis and computation accuracy on the vertical axis.
Focusing only on each method, the computation time increases as the computation accuracy increases,
indicating a trade-off between computation time and accuracy.
Considering the comparison between methods, the lower left it is, the more efficient the calculation.

The LSMC in the figure was measured in the same computing environment by implementing the Least Squares Monte Carlo method.
The calculation method is following.
First, the path of $X_{t}$ is generated by the formula (\ref{formula:G2PPInt1}) in which $U_{t,\tau}$ is replaced by a random number whose variance-covariance is $\bm{\Sigma}(t,\tau)$.
The optimal exercise boundary is estimated by a 2 dimensional second-order polynomial equation
using the plus part of Underlying value and spot rate as an explanatory variable and path cashflow values as explained variable.
Using above boundary, we calculate optimal exercise time and take the average value of the cash flow.
We define this LSMC method with 10000 paths as one set,
calculate the maximum $N_{MC}=3200$ sets and estimate Monte Carlo error by multiplying the standard deviation by $\frac{1}{\sqrt{N_{MC}}}$.

This procedure estimates the optimal exercise bounds with one explanatory variable for a two-factor model, which introduces lower bias into the calculation results.
We intended to remove this bias by using the standard deviation of the mean as the error.

In the result of September 20 2018 data, the accuracy deteriorates rapidly in relation to the calculation time when grid rotation is not used.
However, the deterioration in accuracy is suppressed, and the case with grid rotation is generally superior.
It is shown that the case with FGT is more computationally efficient than the case without FGT and than the LSMC case.
In the case of grid rotation, the deterioration of accuracy is suppressed when using September 20 2018 data and a coarse grid,
which shows the grid rotation improves the computational stability.

From the above experiments, we can conclude that the proposed method using the combination of FGT and grid rotation is stable even when the correlation is close to $-1$.

\subsection{Numerical Example of Expected Exposure}
In this subsection, we experimented the calculation of expected (positive) exposure which appear in CVA and compare difference between rotation or not.

Let $V_{s}$ be the living price at time $s$ of the same 5Y Bermudan swaption as in the previous subsection,
we calculate
\begin{eqnarray}
  \mbox{EPE}(s)&=&\mathbb{E}\left[e^{-\int_{0}^{s}R_{u}du}\max\left\{V_{s},0\right\}_{+}\right].
\end{eqnarray}
The expectation is calculated by Monte Carlo method which generate 10,000 paths of $X_{s}$.
When $V_{s}$ expresses Bermudan swaption in some paths, it is calculated by proposed method with the same settings of $N_{y}=300$ and
as the same way of previous LSMC setting for each day from time 0 to trade maturity and interpolate $V_{s}$ on path $X_{s}$ by bilinear interpolation.
In the case that Bermudan swaption is already exercised until $s$, $V_{s}$ is calculated analytically as fixed or float cashflows.

\begin{figure}[H]
  \begin{tabular}{cc}
    \begin{minipage}[t]{0.45\hsize}
      \centering
      \includegraphics[keepaspectratio, scale=0.4]{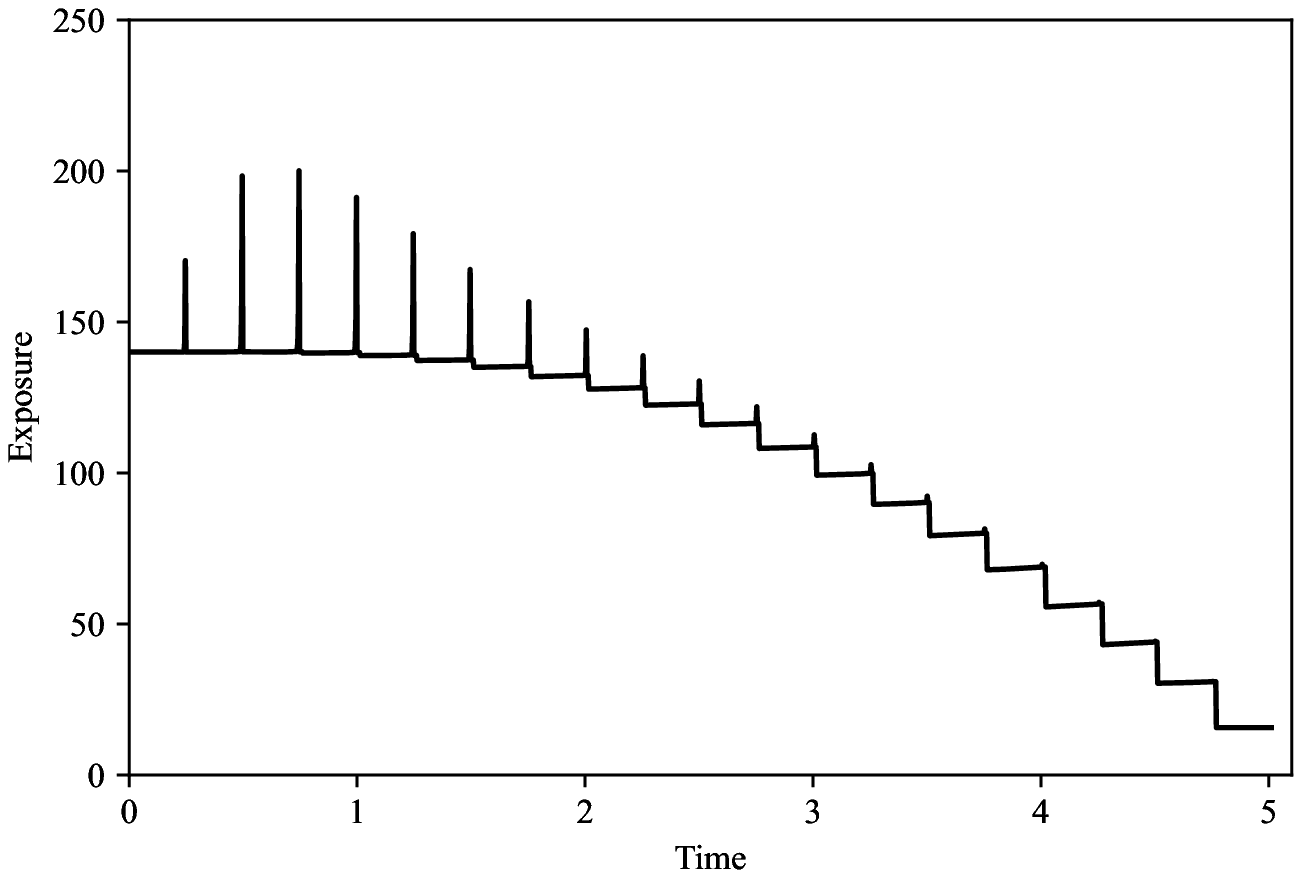}
      \caption{Exposure without grid rotation (Sep 20 2018 data)}
      \label{figure:Ex2_2018EE300NoRotate}
    \end{minipage} &
    \begin{minipage}[t]{0.45\hsize}
      \centering
      \includegraphics[keepaspectratio, scale=0.4]{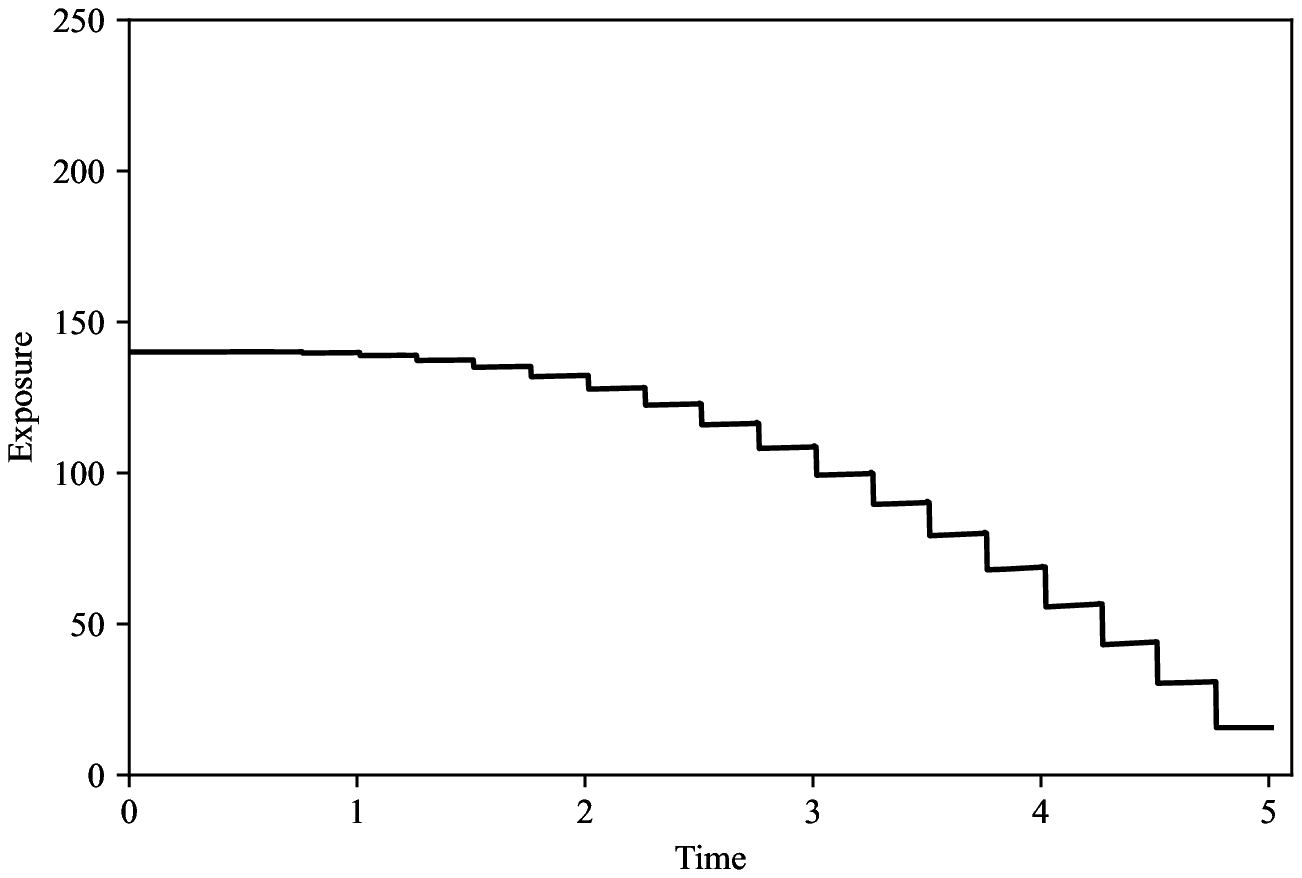}
      \caption{Exposure with grid rotation (Sep 20 2018 data)}
      \label{figure:Ex2_2018EE300Rotate}
    \end{minipage}\\
    \begin{minipage}[t]{0.45\hsize}
      \centering
      \includegraphics[keepaspectratio, scale=0.4]{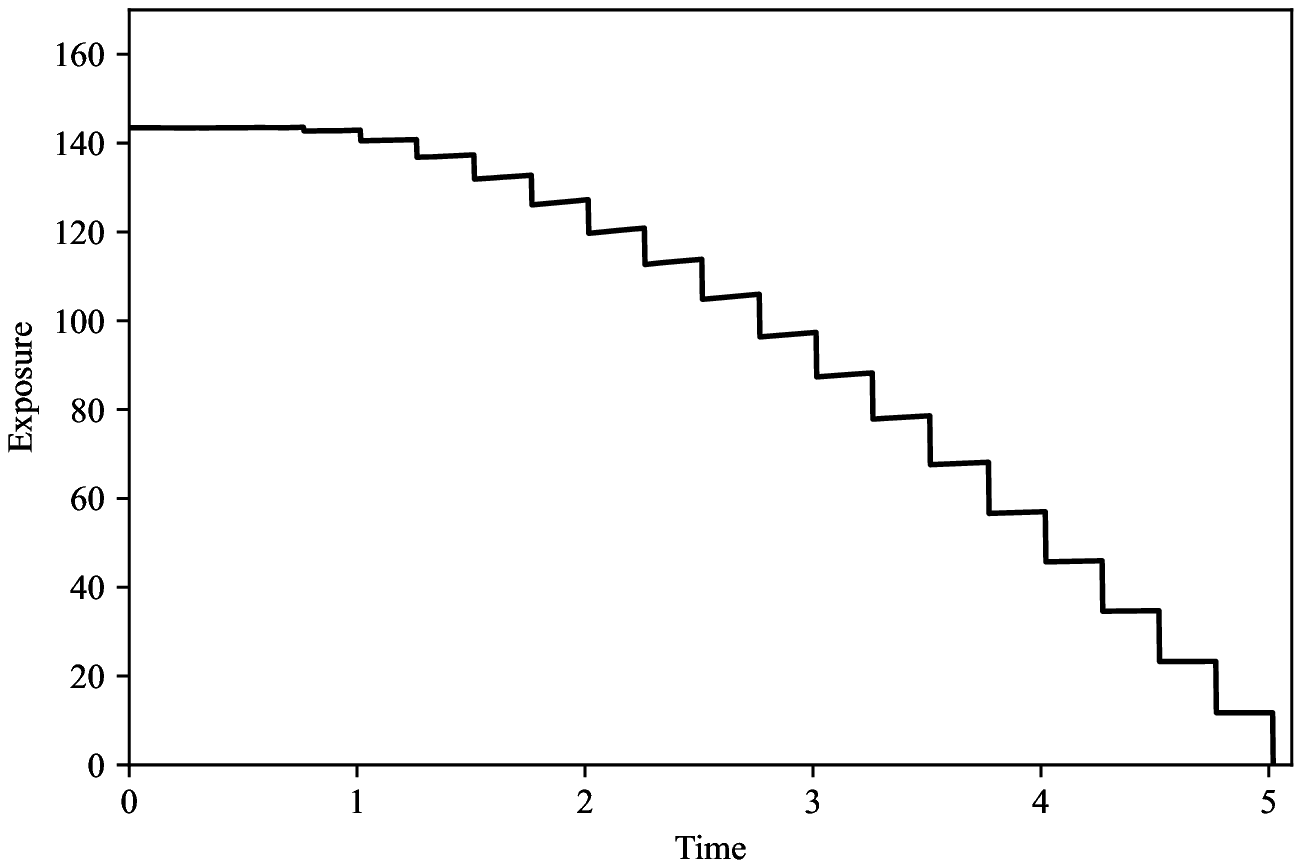}
      \caption{Exposure without grid rotation (Dec 5 2019 data)}
      \label{figure:Ex2_2019EE300NoRotate}
    \end{minipage} &
    \begin{minipage}[t]{0.45\hsize}
      \centering
      \includegraphics[keepaspectratio, scale=0.4]{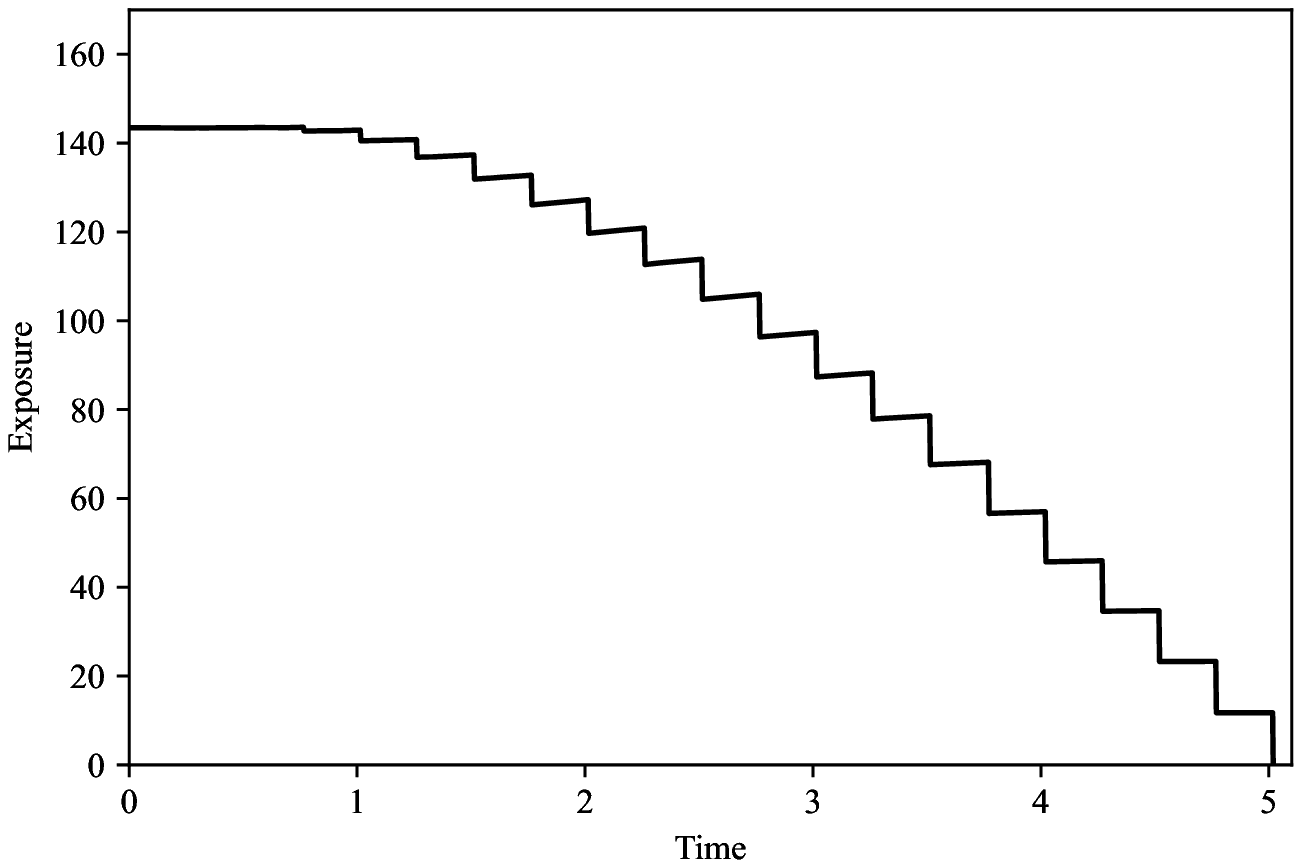}
      \caption{Exposure with grid rotation (Dec 5 2019 data)}
      \label{figure:2019EE300Rotate}
    \end{minipage}
  \end{tabular}
\end{figure}

Figure \ref{figure:Ex2_2018EE300NoRotate} shows a graph of the exposures without applying grid rotation to the September 20 2018 data.
We can see that the values jump, especially just before the exercise which means that the error becomes large when $\tau-t$ is small.
In contrast, no such jumps is found when we use December 5 2019 data or use grid rotation. 

The above results indicate that when the correlation is close to $-1$,
the accuracy of the proposed method without grid rotation tends to be poor,
especially for calculations with short time intervals, and grid rotation is superior from the viewpoint of stability.

\section{Conclusion}
In this study, we developed and evaluated a method for calculating the price of Bermudan swaptions in the two factor Hull-White model by numerical integration combined with the FGT and grid rotation.
In the proposed method, the expected value appearing in the price calculation in the two factor Hull-White model is expressed as the integral of the product of the density function including the coordinate rotation and the payoff and is discretized as a numerical integral by introducing a grid.
For the rotation angle of the coordinate rotation, the optimum value was derived by imposing the condition that the density of grid points is maximized from the viewpoint of the accuracy of numerical integration.
Numerical experiments showed that the grid rotation stabilized the calculation accuracy and improved the calculation speed for market data with correlation close to $-1$ and was superior to the least-squares Monte Carlo method in terms of calculation speed and accuracy.

Furthermore, it is shown that this method can be applied to the calculation of exposures in CVA.
In the numerical experiments, the accuracy deteriorated when the correlation was close to $-1$ and the time of exposures calculation was very close to the time of next exercise, but the stability was improved by the method combining grid rotation.

\appendix
\section{Variables of two factor Hull-White model at constant coefficients}
\label{sec:G2PPFormula}
Consider the case where the coefficients are constant in the two factor Hull-White  model.
\begin{eqnarray*}
  \bm{\kappa}(t)&=&\diag{(\kappa_{0}, \kappa_{1})},\\
  \bm{\sigma}(t)&=&\diag{(\sigma_{0}, \sigma_{1})}.
\end{eqnarray*}
The variables that appear in the equations in the text can be written as follows.
\begin{eqnarray*}
  \chi(a,t)&=&\frac{1-e^{-at}}{a},\\
  \bm{\mu}(t,\tau)&=&
  \diag{(e^{-\kappa_{0}(\tau-t)}, e^{-\kappa_{1}(\tau-t)})},\\
  \bm{\nu}(t,\tau)&=&
  \diag{\left(\chi(\kappa_{0},\tau-t), \chi(\kappa_{1},\tau-t)\right)},\\
  \bm{\Sigma}_{00}(t,\tau)&=&\sigma_{0}^{2}\chi(2\kappa_{0},\tau-t),\\
  \bm{\Sigma}_{10}(t,\tau)&=&\bm{\Sigma}_{01}(t,\tau)=\rho\sigma_{0}\sigma_{1}\chi(\kappa_{0}+\kappa_{1},\tau-t),\\
  \bm{\Sigma}_{11}(t,\tau)&=&\sigma_{1}^{2}\chi(2\kappa_{1},\tau-t),\\
  v(t,\tau)&=&\sigma_{0}^{2}\frac{(\tau-t)-2\chi(\kappa_{0},\tau-t)+\chi(2\kappa_{0},\tau-t)}{\kappa_{0}^{2}}\nonumber\\
  &&+2\rho\sigma_{0}\sigma_{1}\frac{(\tau-t)-\chi(\kappa_{0},\tau-t)-\chi(\kappa_{1},\tau-t)+\chi(\kappa_{0}+\kappa_{1},\tau-t)}{\kappa_{1}^{2}}\nonumber\\
  &&+\sigma_{1}^{2}\frac{(\tau-t)-2\chi(\kappa_{1},\tau-t)+\chi(2\kappa_{1},\tau-t)}{\kappa_{1}^{2}},\\
  \eta_{0}(t,\tau)&=&-\sigma_{0}^{2}\frac{\chi(\kappa_{0},\tau-t)-\chi(2\kappa_{0},\tau-t)}{\kappa_{0}}-\rho\sigma_{0}\sigma_{1}\frac{\chi(\kappa_{0},\tau-t)-\chi(\kappa_{0}+\kappa_{1},\tau-t)}{\kappa_{1}},\\
  \eta_{1}(t,\tau)&=&-\rho\sigma_{0}\sigma_{1}\frac{\chi(\kappa_{1},\tau-t)-\chi(\kappa_{0}+\kappa_{1},\tau-t)}{\kappa_{0}}-\sigma_{1}^{2}\frac{\chi(\kappa_{1},\tau-t)-\chi(2\kappa_{1},\tau-t)}{\kappa_{1}}.
\end{eqnarray*}

\bibliographystyle{plain}
\bibliography{tr}

\end{document}